\documentclass[longauth]{aa}

\usepackage{euclid}
\usepackage{graphicx}
\usepackage[dvipsnames]{xcolor}
\usepackage{multirow}
\usepackage{booktabs}
\usepackage{txfonts}
\usepackage[colorlinks=true,
            linkcolor=blue,
            citecolor=blue,
            urlcolor=blue,
            anchorcolor=blue,
            linkcolor=blue]{hyperref}
\begin{document}

   \title{\Euclid\: Testing photometric selection of emission-line galaxy targets\thanks{This paper is published on behalf of the Euclid Consortium.}}

\renewcommand{\orcid}[1]{} 
\author{M.~S.~Cagliari\orcid{0000-0002-2912-9233}\thanks{\email{marina.cagliari@unimi.it}}\inst{\ref{aff1}}
\and B.~R.~Granett\orcid{0000-0003-2694-9284}\inst{\ref{aff2}}
\and L.~Guzzo\orcid{0000-0001-8264-5192}\inst{\ref{aff1},\ref{aff2},\ref{aff3}}
\and M.~Bethermin\orcid{0000-0002-3915-2015}\inst{\ref{aff4},\ref{aff5}}
\and M.~Bolzonella\orcid{0000-0003-3278-4607}\inst{\ref{aff6}}
\and S.~de~la~Torre\inst{\ref{aff5}}
\and P.~Monaco\orcid{0000-0003-2083-7564}\inst{\ref{aff7},\ref{aff8},\ref{aff9},\ref{aff10}}
\and M.~Moresco\orcid{0000-0002-7616-7136}\inst{\ref{aff11},\ref{aff6}}
\and W.~J.~Percival\orcid{0000-0002-0644-5727}\inst{\ref{aff12},\ref{aff13},\ref{aff14}}
\and C.~Scarlata\orcid{0000-0002-9136-8876}\inst{\ref{aff15}}
\and Y.~Wang\orcid{0000-0002-4749-2984}\inst{\ref{aff16}}
\and M.~Ezziati\orcid{0009-0003-6065-1585}\inst{\ref{aff5}}
\and O.~Ilbert\orcid{0000-0002-7303-4397}\inst{\ref{aff5}}
\and V.~Le~Brun\inst{\ref{aff5}}
\and A.~Amara\inst{\ref{aff17}}
\and S.~Andreon\orcid{0000-0002-2041-8784}\inst{\ref{aff2}}
\and N.~Auricchio\orcid{0000-0003-4444-8651}\inst{\ref{aff6}}
\and M.~Baldi\orcid{0000-0003-4145-1943}\inst{\ref{aff18},\ref{aff6},\ref{aff19}}
\and S.~Bardelli\orcid{0000-0002-8900-0298}\inst{\ref{aff6}}
\and R.~Bender\orcid{0000-0001-7179-0626}\inst{\ref{aff20},\ref{aff21}}
\and C.~Bodendorf\inst{\ref{aff20}}
\and E.~Branchini\orcid{0000-0002-0808-6908}\inst{\ref{aff22},\ref{aff23},\ref{aff2}}
\and M.~Brescia\orcid{0000-0001-9506-5680}\inst{\ref{aff24},\ref{aff25},\ref{aff26}}
\and J.~Brinchmann\orcid{0000-0003-4359-8797}\inst{\ref{aff27}}
\and S.~Camera\orcid{0000-0003-3399-3574}\inst{\ref{aff28},\ref{aff29},\ref{aff30}}
\and V.~Capobianco\orcid{0000-0002-3309-7692}\inst{\ref{aff30}}
\and C.~Carbone\orcid{0000-0003-0125-3563}\inst{\ref{aff31}}
\and J.~Carretero\orcid{0000-0002-3130-0204}\inst{\ref{aff32},\ref{aff33}}
\and S.~Casas\orcid{0000-0002-4751-5138}\inst{\ref{aff34}}
\and M.~Castellano\orcid{0000-0001-9875-8263}\inst{\ref{aff35}}
\and S.~Cavuoti\orcid{0000-0002-3787-4196}\inst{\ref{aff25},\ref{aff26}}
\and A.~Cimatti\inst{\ref{aff36}}
\and G.~Congedo\orcid{0000-0003-2508-0046}\inst{\ref{aff37}}
\and C.~J.~Conselice\inst{\ref{aff38}}
\and L.~Conversi\orcid{0000-0002-6710-8476}\inst{\ref{aff39},\ref{aff40}}
\and Y.~Copin\orcid{0000-0002-5317-7518}\inst{\ref{aff41}}
\and L.~Corcione\orcid{0000-0002-6497-5881}\inst{\ref{aff30}}
\and F.~Courbin\orcid{0000-0003-0758-6510}\inst{\ref{aff42}}
\and H.~M.~Courtois\orcid{0000-0003-0509-1776}\inst{\ref{aff43}}
\and A.~Da~Silva\orcid{0000-0002-6385-1609}\inst{\ref{aff44},\ref{aff45}}
\and H.~Degaudenzi\orcid{0000-0002-5887-6799}\inst{\ref{aff46}}
\and A.~M.~Di~Giorgio\orcid{0000-0002-4767-2360}\inst{\ref{aff47}}
\and J.~Dinis\inst{\ref{aff45},\ref{aff44}}
\and F.~Dubath\orcid{0000-0002-6533-2810}\inst{\ref{aff46}}
\and C.~A.~J.~Duncan\inst{\ref{aff38},\ref{aff48}}
\and X.~Dupac\inst{\ref{aff40}}
\and S.~Dusini\orcid{0000-0002-1128-0664}\inst{\ref{aff49}}
\and A.~Ealet\inst{\ref{aff50}}
\and M.~Farina\orcid{0000-0002-3089-7846}\inst{\ref{aff47}}
\and S.~Farrens\orcid{0000-0002-9594-9387}\inst{\ref{aff51}}
\and S.~Ferriol\inst{\ref{aff41}}
\and S.~Fotopoulou\orcid{0000-0002-9686-254X}\inst{\ref{aff52}}
\and M.~Frailis\orcid{0000-0002-7400-2135}\inst{\ref{aff8}}
\and E.~Franceschi\orcid{0000-0002-0585-6591}\inst{\ref{aff6}}
\and S.~Galeotta\orcid{0000-0002-3748-5115}\inst{\ref{aff8}}
\and B.~Gillis\orcid{0000-0002-4478-1270}\inst{\ref{aff37}}
\and C.~Giocoli\orcid{0000-0002-9590-7961}\inst{\ref{aff6},\ref{aff53}}
\and A.~Grazian\orcid{0000-0002-5688-0663}\inst{\ref{aff54}}
\and F.~Grupp\inst{\ref{aff20},\ref{aff55}}
\and S.~V.~H.~Haugan\orcid{0000-0001-9648-7260}\inst{\ref{aff56}}
\and H.~Hoekstra\orcid{0000-0002-0641-3231}\inst{\ref{aff57}}
\and I.~Hook\inst{\ref{aff58}}
\and F.~Hormuth\inst{\ref{aff59}}
\and A.~Hornstrup\orcid{0000-0002-3363-0936}\inst{\ref{aff60},\ref{aff61}}
\and K.~Jahnke\orcid{0000-0003-3804-2137}\inst{\ref{aff62}}
\and E.~Keih\"anen\orcid{0000-0003-1804-7715}\inst{\ref{aff63}}
\and S.~Kermiche\orcid{0000-0002-0302-5735}\inst{\ref{aff64}}
\and A.~Kiessling\orcid{0000-0002-2590-1273}\inst{\ref{aff65}}
\and M.~Kilbinger\orcid{0000-0001-9513-7138}\inst{\ref{aff66}}
\and B.~Kubik\inst{\ref{aff41}}
\and M.~K\"ummel\orcid{0000-0003-2791-2117}\inst{\ref{aff21}}
\and M.~Kunz\orcid{0000-0002-3052-7394}\inst{\ref{aff67}}
\and H.~Kurki-Suonio\orcid{0000-0002-4618-3063}\inst{\ref{aff68},\ref{aff69}}
\and S.~Ligori\orcid{0000-0003-4172-4606}\inst{\ref{aff30}}
\and P.~B.~Lilje\orcid{0000-0003-4324-7794}\inst{\ref{aff56}}
\and V.~Lindholm\orcid{0000-0003-2317-5471}\inst{\ref{aff68},\ref{aff69}}
\and I.~Lloro\inst{\ref{aff70}}
\and D.~Maino\inst{\ref{aff1},\ref{aff31},\ref{aff3}}
\and E.~Maiorano\orcid{0000-0003-2593-4355}\inst{\ref{aff6}}
\and O.~Mansutti\orcid{0000-0001-5758-4658}\inst{\ref{aff8}}
\and O.~Marggraf\orcid{0000-0001-7242-3852}\inst{\ref{aff71}}
\and K.~Markovic\orcid{0000-0001-6764-073X}\inst{\ref{aff65}}
\and N.~Martinet\orcid{0000-0003-2786-7790}\inst{\ref{aff5}}
\and F.~Marulli\orcid{0000-0002-8850-0303}\inst{\ref{aff11},\ref{aff6},\ref{aff19}}
\and R.~Massey\orcid{0000-0002-6085-3780}\inst{\ref{aff72}}
\and S.~Maurogordato\inst{\ref{aff73}}
\and H.~J.~McCracken\orcid{0000-0002-9489-7765}\inst{\ref{aff74}}
\and E.~Medinaceli\orcid{0000-0002-4040-7783}\inst{\ref{aff6}}
\and S.~Mei\orcid{0000-0002-2849-559X}\inst{\ref{aff75}}
\and Y.~Mellier\inst{\ref{aff76},\ref{aff74}}
\and M.~Meneghetti\orcid{0000-0003-1225-7084}\inst{\ref{aff6},\ref{aff19}}
\and E.~Merlin\orcid{0000-0001-6870-8900}\inst{\ref{aff35}}
\and G.~Meylan\inst{\ref{aff42}}
\and L.~Moscardini\orcid{0000-0002-3473-6716}\inst{\ref{aff11},\ref{aff6},\ref{aff19}}
\and E.~Munari\orcid{0000-0002-1751-5946}\inst{\ref{aff8}}
\and R.~C.~Nichol\inst{\ref{aff17}}
\and S.-M.~Niemi\inst{\ref{aff77}}
\and C.~Padilla\orcid{0000-0001-7951-0166}\inst{\ref{aff32}}
\and S.~Paltani\inst{\ref{aff46}}
\and F.~Pasian\inst{\ref{aff8}}
\and K.~Pedersen\inst{\ref{aff78}}
\and V.~Pettorino\inst{\ref{aff79}}
\and S.~Pires\orcid{0000-0002-0249-2104}\inst{\ref{aff51}}
\and G.~Polenta\orcid{0000-0003-4067-9196}\inst{\ref{aff80}}
\and M.~Poncet\inst{\ref{aff81}}
\and L.~A.~Popa\inst{\ref{aff82}}
\and L.~Pozzetti\orcid{0000-0001-7085-0412}\inst{\ref{aff6}}
\and F.~Raison\orcid{0000-0002-7819-6918}\inst{\ref{aff20}}
\and R.~Rebolo\inst{\ref{aff83},\ref{aff84}}
\and A.~Renzi\orcid{0000-0001-9856-1970}\inst{\ref{aff85},\ref{aff49}}
\and J.~Rhodes\inst{\ref{aff65}}
\and G.~Riccio\inst{\ref{aff25}}
\and E.~Romelli\orcid{0000-0003-3069-9222}\inst{\ref{aff8}}
\and M.~Roncarelli\orcid{0000-0001-9587-7822}\inst{\ref{aff6}}
\and E.~Rossetti\inst{\ref{aff18}}
\and R.~Saglia\orcid{0000-0003-0378-7032}\inst{\ref{aff21},\ref{aff20}}
\and D.~Sapone\orcid{0000-0001-7089-4503}\inst{\ref{aff86}}
\and B.~Sartoris\inst{\ref{aff21},\ref{aff8}}
\and P.~Schneider\orcid{0000-0001-8561-2679}\inst{\ref{aff71}}
\and M.~Scodeggio\inst{\ref{aff31}}
\and A.~Secroun\orcid{0000-0003-0505-3710}\inst{\ref{aff64}}
\and G.~Seidel\orcid{0000-0003-2907-353X}\inst{\ref{aff62}}
\and M.~Seiffert\orcid{0000-0002-7536-9393}\inst{\ref{aff65}}
\and S.~Serrano\orcid{0000-0002-0211-2861}\inst{\ref{aff87},\ref{aff88},\ref{aff89}}
\and C.~Sirignano\orcid{0000-0002-0995-7146}\inst{\ref{aff85},\ref{aff49}}
\and G.~Sirri\orcid{0000-0003-2626-2853}\inst{\ref{aff19}}
\and J.~Skottfelt\orcid{0000-0003-1310-8283}\inst{\ref{aff90}}
\and L.~Stanco\orcid{0000-0002-9706-5104}\inst{\ref{aff49}}
\and C.~Surace\inst{\ref{aff5}}
\and A.~N.~Taylor\inst{\ref{aff37}}
\and H.~I.~Teplitz\orcid{0000-0002-7064-5424}\inst{\ref{aff16}}
\and I.~Tereno\inst{\ref{aff44},\ref{aff91}}
\and R.~Toledo-Moreo\orcid{0000-0002-2997-4859}\inst{\ref{aff92}}
\and F.~Torradeflot\orcid{0000-0003-1160-1517}\inst{\ref{aff33},\ref{aff93}}
\and I.~Tutusaus\orcid{0000-0002-3199-0399}\inst{\ref{aff94}}
\and E.~A.~Valentijn\inst{\ref{aff95}}
\and L.~Valenziano\orcid{0000-0002-1170-0104}\inst{\ref{aff6},\ref{aff96}}
\and T.~Vassallo\orcid{0000-0001-6512-6358}\inst{\ref{aff21},\ref{aff8}}
\and A.~Veropalumbo\orcid{0000-0003-2387-1194}\inst{\ref{aff2},\ref{aff23}}
\and J.~Weller\orcid{0000-0002-8282-2010}\inst{\ref{aff21},\ref{aff20}}
\and G.~Zamorani\orcid{0000-0002-2318-301X}\inst{\ref{aff6}}
\and J.~Zoubian\inst{\ref{aff64}}
\and E.~Zucca\orcid{0000-0002-5845-8132}\inst{\ref{aff6}}
\and C.~Burigana\orcid{0000-0002-3005-5796}\inst{\ref{aff97},\ref{aff96}}
\and V.~Scottez\inst{\ref{aff76},\ref{aff98}}
\and M.~Viel\orcid{0000-0002-2642-5707}\inst{\ref{aff10},\ref{aff8},\ref{aff99},\ref{aff9},\ref{aff100}}
\and L.~Bisigello\orcid{0000-0003-0492-4924}\inst{\ref{aff97},\ref{aff85}}}
										   
\institute{Dipartimento di Fisica "Aldo Pontremoli", Universit\`a degli Studi di Milano, Via Celoria 16, 20133 Milano, Italy\label{aff1}
\and
INAF-Osservatorio Astronomico di Brera, Via Brera 28, 20122 Milano, Italy\label{aff2}
\and
INFN-Sezione di Milano, Via Celoria 16, 20133 Milano, Italy\label{aff3}
\and
Universit\'e de Strasbourg, CNRS, Observatoire astronomique de Strasbourg, UMR 7550, 67000 Strasbourg, France\label{aff4}
\and
Aix-Marseille Universit\'e, CNRS, CNES, LAM, Marseille, France\label{aff5}
\and
INAF-Osservatorio di Astrofisica e Scienza dello Spazio di Bologna, Via Piero Gobetti 93/3, 40129 Bologna, Italy\label{aff6}
\and
Dipartimento di Fisica - Sezione di Astronomia, Universit\`a di Trieste, Via Tiepolo 11, 34131 Trieste, Italy\label{aff7}
\and
INAF-Osservatorio Astronomico di Trieste, Via G. B. Tiepolo 11, 34143 Trieste, Italy\label{aff8}
\and
INFN, Sezione di Trieste, Via Valerio 2, 34127 Trieste TS, Italy\label{aff9}
\and
IFPU, Institute for Fundamental Physics of the Universe, via Beirut 2, 34151 Trieste, Italy\label{aff10}
\and
Dipartimento di Fisica e Astronomia "Augusto Righi" - Alma Mater Studiorum Universit\`a di Bologna, via Piero Gobetti 93/2, 40129 Bologna, Italy\label{aff11}
\and
Centre for Astrophysics, University of Waterloo, Waterloo, Ontario N2L 3G1, Canada\label{aff12}
\and
Department of Physics and Astronomy, University of Waterloo, Waterloo, Ontario N2L 3G1, Canada\label{aff13}
\and
Perimeter Institute for Theoretical Physics, Waterloo, Ontario N2L 2Y5, Canada\label{aff14}
\and
Minnesota Institute for Astrophysics, University of Minnesota, 116 Church St SE, Minneapolis, MN 55455, USA\label{aff15}
\and
Infrared Processing and Analysis Center, California Institute of Technology, Pasadena, CA 91125, USA\label{aff16}
\and
School of Mathematics and Physics, University of Surrey, Guildford, Surrey, GU2 7XH, UK\label{aff17}
\and
Dipartimento di Fisica e Astronomia, Universit\`a di Bologna, Via Gobetti 93/2, 40129 Bologna, Italy\label{aff18}
\and
INFN-Sezione di Bologna, Viale Berti Pichat 6/2, 40127 Bologna, Italy\label{aff19}
\and
Max Planck Institute for Extraterrestrial Physics, Giessenbachstr. 1, 85748 Garching, Germany\label{aff20}
\and
Universit\"ats-Sternwarte M\"unchen, Fakult\"at f\"ur Physik, Ludwig-Maximilians-Universit\"at M\"unchen, Scheinerstrasse 1, 81679 M\"unchen, Germany\label{aff21}
\and
Dipartimento di Fisica, Universit\`a di Genova, Via Dodecaneso 33, 16146, Genova, Italy\label{aff22}
\and
INFN-Sezione di Genova, Via Dodecaneso 33, 16146, Genova, Italy\label{aff23}
\and
Department of Physics "E. Pancini", University Federico II, Via Cinthia 6, 80126, Napoli, Italy\label{aff24}
\and
INAF-Osservatorio Astronomico di Capodimonte, Via Moiariello 16, 80131 Napoli, Italy\label{aff25}
\and
INFN section of Naples, Via Cinthia 6, 80126, Napoli, Italy\label{aff26}
\and
Instituto de Astrof\'isica e Ci\^encias do Espa\c{c}o, Universidade do Porto, CAUP, Rua das Estrelas, PT4150-762 Porto, Portugal\label{aff27}
\and
Dipartimento di Fisica, Universit\`a degli Studi di Torino, Via P. Giuria 1, 10125 Torino, Italy\label{aff28}
\and
INFN-Sezione di Torino, Via P. Giuria 1, 10125 Torino, Italy\label{aff29}
\and
INAF-Osservatorio Astrofisico di Torino, Via Osservatorio 20, 10025 Pino Torinese (TO), Italy\label{aff30}
\and
INAF-IASF Milano, Via Alfonso Corti 12, 20133 Milano, Italy\label{aff31}
\and
Institut de F\'{i}sica d'Altes Energies (IFAE), The Barcelona Institute of Science and Technology, Campus UAB, 08193 Bellaterra (Barcelona), Spain\label{aff32}
\and
Port d'Informaci\'{o} Cient\'{i}fica, Campus UAB, C. Albareda s/n, 08193 Bellaterra (Barcelona), Spain\label{aff33}
\and
Institute for Theoretical Particle Physics and Cosmology (TTK), RWTH Aachen University, 52056 Aachen, Germany\label{aff34}
\and
INAF-Osservatorio Astronomico di Roma, Via Frascati 33, 00078 Monteporzio Catone, Italy\label{aff35}
\and
Dipartimento di Fisica e Astronomia "Augusto Righi" - Alma Mater Studiorum Universit\`a di Bologna, Viale Berti Pichat 6/2, 40127 Bologna, Italy\label{aff36}
\and
Institute for Astronomy, University of Edinburgh, Royal Observatory, Blackford Hill, Edinburgh EH9 3HJ, UK\label{aff37}
\and
Jodrell Bank Centre for Astrophysics, Department of Physics and Astronomy, University of Manchester, Oxford Road, Manchester M13 9PL, UK\label{aff38}
\and
European Space Agency/ESRIN, Largo Galileo Galilei 1, 00044 Frascati, Roma, Italy\label{aff39}
\and
ESAC/ESA, Camino Bajo del Castillo, s/n., Urb. Villafranca del Castillo, 28692 Villanueva de la Ca\~nada, Madrid, Spain\label{aff40}
\and
University of Lyon, Univ Claude Bernard Lyon 1, CNRS/IN2P3, IP2I Lyon, UMR 5822, 69622 Villeurbanne, France\label{aff41}
\and
Institute of Physics, Laboratory of Astrophysics, Ecole Polytechnique F\'ed\'erale de Lausanne (EPFL), Observatoire de Sauverny, 1290 Versoix, Switzerland\label{aff42}
\and
UCB Lyon 1, CNRS/IN2P3, IUF, IP2I Lyon, 4 rue Enrico Fermi, 69622 Villeurbanne, France\label{aff43}
\and
Departamento de F\'isica, Faculdade de Ci\^encias, Universidade de Lisboa, Edif\'icio C8, Campo Grande, PT1749-016 Lisboa, Portugal\label{aff44}
\and
Instituto de Astrof\'isica e Ci\^encias do Espa\c{c}o, Faculdade de Ci\^encias, Universidade de Lisboa, Campo Grande, 1749-016 Lisboa, Portugal\label{aff45}
\and
Department of Astronomy, University of Geneva, ch. d'Ecogia 16, 1290 Versoix, Switzerland\label{aff46}
\and
INAF-Istituto di Astrofisica e Planetologia Spaziali, via del Fosso del Cavaliere, 100, 00100 Roma, Italy\label{aff47}
\and
Department of Physics, Oxford University, Keble Road, Oxford OX1 3RH, UK\label{aff48}
\and
INFN-Padova, Via Marzolo 8, 35131 Padova, Italy\label{aff49}
\and
Univ Claude Bernard Lyon 1, CNRS, IP2I Lyon, UMR 5822, 69622 Villeurbanne, France\label{aff50}
\and
Universit\'e Paris-Saclay, Universit\'e Paris Cit\'e, CEA, CNRS, AIM, 91191, Gif-sur-Yvette, France\label{aff51}
\and
School of Physics, HH Wills Physics Laboratory, University of Bristol, Tyndall Avenue, Bristol, BS8 1TL, UK\label{aff52}
\and
Istituto Nazionale di Fisica Nucleare, Sezione di Bologna, Via Irnerio 46, 40126 Bologna, Italy\label{aff53}
\and
INAF-Osservatorio Astronomico di Padova, Via dell'Osservatorio 5, 35122 Padova, Italy\label{aff54}
\and
University Observatory, Faculty of Physics, Ludwig-Maximilians-Universit{\"a}t, Scheinerstr. 1, 81679 Munich, Germany\label{aff55}
\and
Institute of Theoretical Astrophysics, University of Oslo, P.O. Box 1029 Blindern, 0315 Oslo, Norway\label{aff56}
\and
Leiden Observatory, Leiden University, Niels Bohrweg 2, 2333 CA Leiden, The Netherlands\label{aff57}
\and
Department of Physics, Lancaster University, Lancaster, LA1 4YB, UK\label{aff58}
\and
von Hoerner \& Sulger GmbH, Schlo{\ss}Platz 8, 68723 Schwetzingen, Germany\label{aff59}
\and
Technical University of Denmark, Elektrovej 327, 2800 Kgs. Lyngby, Denmark\label{aff60}
\and
Cosmic Dawn Center (DAWN), Denmark\label{aff61}
\and
Max-Planck-Institut f\"ur Astronomie, K\"onigstuhl 17, 69117 Heidelberg, Germany\label{aff62}
\and
Department of Physics and Helsinki Institute of Physics, Gustaf H\"allstr\"omin katu 2, 00014 University of Helsinki, Finland\label{aff63}
\and
Aix-Marseille Universit\'e, CNRS/IN2P3, CPPM, Marseille, France\label{aff64}
\and
Jet Propulsion Laboratory, California Institute of Technology, 4800 Oak Grove Drive, Pasadena, CA, 91109, USA\label{aff65}
\and
AIM, CEA, CNRS, Universit\'{e} Paris-Saclay, Universit\'{e} de Paris, 91191 Gif-sur-Yvette, France\label{aff66}
\and
Universit\'e de Gen\`eve, D\'epartement de Physique Th\'eorique and Centre for Astroparticle Physics, 24 quai Ernest-Ansermet, CH-1211 Gen\`eve 4, Switzerland\label{aff67}
\and
Department of Physics, P.O. Box 64, 00014 University of Helsinki, Finland\label{aff68}
\and
Helsinki Institute of Physics, Gustaf H{\"a}llstr{\"o}min katu 2, University of Helsinki, Helsinki, Finland\label{aff69}
\and
NOVA optical infrared instrumentation group at ASTRON, Oude Hoogeveensedijk 4, 7991PD, Dwingeloo, The Netherlands\label{aff70}
\and
Universit\"at Bonn, Argelander-Institut f\"ur Astronomie, Auf dem H\"ugel 71, 53121 Bonn, Germany\label{aff71}
\and
Department of Physics, Institute for Computational Cosmology, Durham University, South Road, DH1 3LE, UK\label{aff72}
\and
Universit\'e C\^{o}te d'Azur, Observatoire de la C\^{o}te d'Azur, CNRS, Laboratoire Lagrange, Bd de l'Observatoire, CS 34229, 06304 Nice cedex 4, France\label{aff73}
\and
Institut d'Astrophysique de Paris, UMR 7095, CNRS, and Sorbonne Universit\'e, 98 bis boulevard Arago, 75014 Paris, France\label{aff74}
\and
Universit\'e Paris Cit\'e, CNRS, Astroparticule et Cosmologie, 75013 Paris, France\label{aff75}
\and
Institut d'Astrophysique de Paris, 98bis Boulevard Arago, 75014, Paris, France\label{aff76}
\and
European Space Agency/ESTEC, Keplerlaan 1, 2201 AZ Noordwijk, The Netherlands\label{aff77}
\and
Department of Physics and Astronomy, University of Aarhus, Ny Munkegade 120, DK-8000 Aarhus C, Denmark\label{aff78}
\and
Universit\'e Paris-Saclay, Universit\'e Paris Cit\'e, CEA, CNRS, Astrophysique, Instrumentation et Mod\'elisation Paris-Saclay, 91191 Gif-sur-Yvette, France\label{aff79}
\and
Space Science Data Center, Italian Space Agency, via del Politecnico snc, 00133 Roma, Italy\label{aff80}
\and
Centre National d'Etudes Spatiales -- Centre spatial de Toulouse, 18 avenue Edouard Belin, 31401 Toulouse Cedex 9, France\label{aff81}
\and
Institute of Space Science, Str. Atomistilor, nr. 409 M\u{a}gurele, Ilfov, 077125, Romania\label{aff82}
\and
Instituto de Astrof\'isica de Canarias, Calle V\'ia L\'actea s/n, 38204, San Crist\'obal de La Laguna, Tenerife, Spain\label{aff83}
\and
Departamento de Astrof\'isica, Universidad de La Laguna, 38206, La Laguna, Tenerife, Spain\label{aff84}
\and
Dipartimento di Fisica e Astronomia "G. Galilei", Universit\`a di Padova, Via Marzolo 8, 35131 Padova, Italy\label{aff85}
\and
Departamento de F\'isica, FCFM, Universidad de Chile, Blanco Encalada 2008, Santiago, Chile\label{aff86}
\and
Institut d'Estudis Espacials de Catalunya (IEEC), Carrer Gran Capit\'a 2-4, 08034 Barcelona, Spain\label{aff87}
\and
Institute of Space Sciences (ICE, CSIC), Campus UAB, Carrer de Can Magrans, s/n, 08193 Barcelona, Spain\label{aff88}
\and
Satlantis, University Science Park, Sede Bld 48940, Leioa-Bilbao, Spain\label{aff89}
\and
Centre for Electronic Imaging, Open University, Walton Hall, Milton Keynes, MK7~6AA, UK\label{aff90}
\and
Instituto de Astrof\'isica e Ci\^encias do Espa\c{c}o, Faculdade de Ci\^encias, Universidade de Lisboa, Tapada da Ajuda, 1349-018 Lisboa, Portugal\label{aff91}
\and
Universidad Polit\'ecnica de Cartagena, Departamento de Electr\'onica y Tecnolog\'ia de Computadoras,  Plaza del Hospital 1, 30202 Cartagena, Spain\label{aff92}
\and
Centro de Investigaciones Energ\'eticas, Medioambientales y Tecnol\'ogicas (CIEMAT), Avenida Complutense 40, 28040 Madrid, Spain\label{aff93}
\and
Institut de Recherche en Astrophysique et Plan\'etologie (IRAP), Universit\'e de Toulouse, CNRS, UPS, CNES, 14 Av. Edouard Belin, 31400 Toulouse, France\label{aff94}
\and
Kapteyn Astronomical Institute, University of Groningen, PO Box 800, 9700 AV Groningen, The Netherlands\label{aff95}
\and
INFN-Bologna, Via Irnerio 46, 40126 Bologna, Italy\label{aff96}
\and
INAF, Istituto di Radioastronomia, Via Piero Gobetti 101, 40129 Bologna, Italy\label{aff97}
\and
Junia, EPA department, 41 Bd Vauban, 59800 Lille, France\label{aff98}
\and
SISSA, International School for Advanced Studies, Via Bonomea 265, 34136 Trieste TS, Italy\label{aff99}
\and
ICSC - Centro Nazionale di Ricerca in High Performance Computing, Big Data e Quantum Computing, Via Magnanelli 2, Bologna, Italy\label{aff100}}    

   \date{}


\abstract{Multi-object spectroscopic galaxy surveys typically make use of photometric and colour criteria to select their targets. That is not the case of \Euclid, which will use the NISP slitless spectrograph to record spectra for every source over its field of view. Slitless spectroscopy has the advantage of avoiding defining a priori a specific galaxy sample, but at the price of making the selection function harder to quantify. In its Wide Survey, \Euclid aims at building robust statistical samples of emission-line galaxies with fluxes brighter than $\num{2e-16}\, \unit{erg.s^{-1}.cm^{-2}}$, using the H$\alpha$-$\left[\ion{N}{ii}\right]$ complex to measure redshifts within the range $[0.9, 1.8]$. Given the expected signal-to-noise ratio of NISP spectra, at such faint fluxes a significant contamination by wrongly measured redshifts is expected, either due to misidentification of other emission lines, or to noise fluctuations mistaken as such, with the consequence of reducing the purity of the final samples.
This can be significantly ameliorated by exploiting the extensive \Euclid photometric information to identify emission-line galaxies over the redshift range of interest. 
Beyond classical multi-band selections in colour space, machine learning techniques provide novel tools to perform this task. Here, we compare and quantify the performance of six such classification algorithms in achieving this goal. We consider the case when only the \Euclid photometric and morphological measurements are used, and when these are supplemented by the extensive set of ancillary ground-based photometric data, which are part of the overall \Euclid scientific strategy to perform lensing tomography. The classifiers are trained and tested on two mock galaxy samples, the EL-COSMOS and Euclid Flagship2 catalogues. The best performance is obtained from either a dense neural network or a support vector classifier, with comparable results in terms of the adopted metrics. When training on \Euclid on-board photometry alone, these are able to remove $87\%$ of the sources that are fainter than the nominal flux limit or lie outside the $0.9<z<1.8$ redshift range, a figure that increases to $97\%$ when ground-based photometry is included. These results show how by using the photometric information available to \Euclid it will be possible to efficiently identify and discard spurious interlopers, allowing us to build robust spectroscopic samples for cosmological investigations.
}


   \maketitle
%

\section{Introduction} \label{sec:intro}
\begin{figure*}
    \centering
    \includegraphics[width=.8\textwidth]{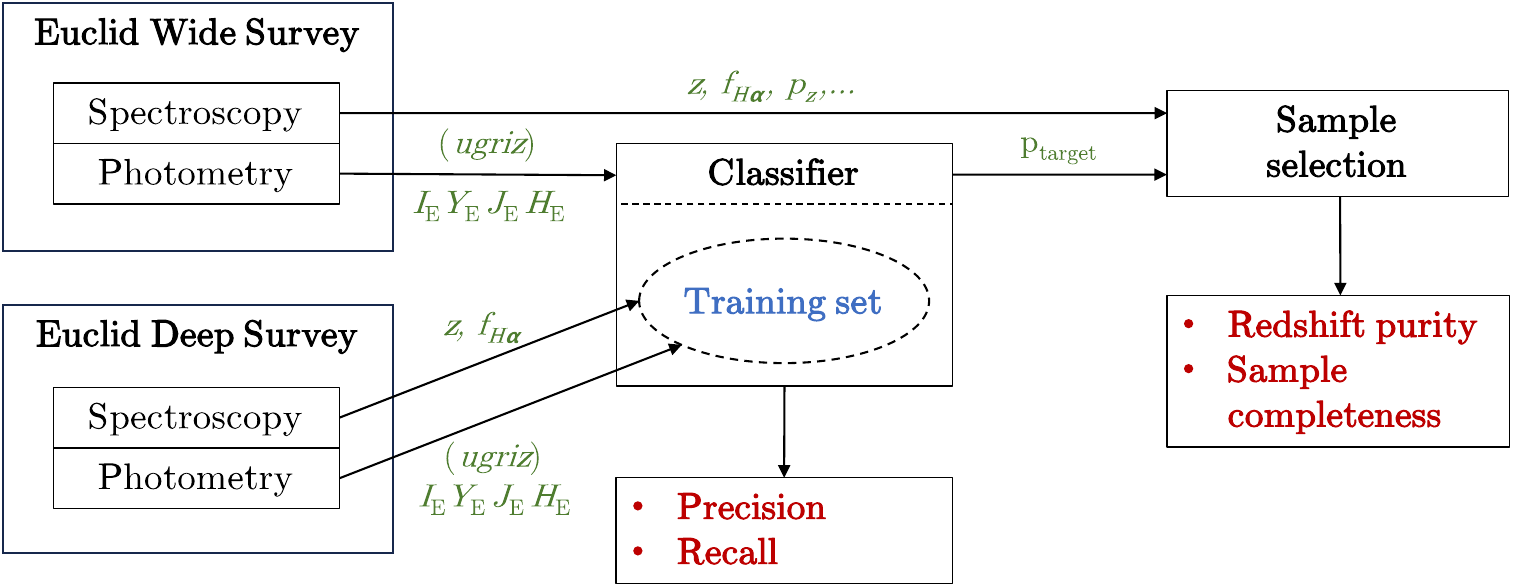}
    \caption{Schematic description of the spectroscopic sample selection pipeline. The flowchart shows where a photometric target selection would be inserted in the spectroscopic selection pipeline. The photometric classifier performance is quantified by its precision and recall (defined in Sect.~\ref{sec:comparison_metrics}), while the final spectroscopic sample is characterised by the redshift purity and sample completeness.}
    \label{fig:flowchart}
\end{figure*}
The ESA \Euclid mission will carry out an imaging and spectroscopic survey over one third of the sky \citep{euclid}. The imaging channel will enable measurements of cosmic shear providing a tomographic view of the matter distribution, while the spectroscopic redshift survey will map the large-scale structure in three dimensions. Jointly, the two probes will yield unprecedented constraints on the cosmological model \citep{2020A&A...642A.191E}.

The \Euclid near-infrared spectrograph and photometer \citep[NISP;][]{Maciaszek:2022} has three broadband filters for imaging, $\YE$, $\JE$, and $\HE$ \citep{Schirmer2022} and a set of grisms for spectroscopy, while the visual instrument \citep[VIS;][]{Cropper:2016} images through a single broad pass band, $\IE$, spanning the range $[530,920]\,\unit{nm}$, with high spatial resolution of $0.1 \,  \unit{arcsec/pixel}$. Jointly, these two instruments will carry out the Euclid Wide and Deep Surveys \citep{EuclidWide}.
The NISP instrument operates as a slitless spectrograph, to record the dispersed light of all sources in the field of view to a nominal emission-line flux limit of $\num{2e-16}\, \unit{erg.s^{-1}.cm^{-2}}$, which corresponds to a $3.5 \sigma$ detection of a $0.5 \, \text{arcsec}$ diameter source in the Wide survey as designed. The use of slitless spectroscopy makes the spectroscopic survey highly efficient, since individual sources do not need to be targeted; however, reliable redshift measurements will only be secured for a fraction of the galaxies that are detected photometrically. The Wide Survey will detect the most luminous H$\alpha$ emitters over the redshift range $0.9 < z < 1.8$, with typical broadband flux corresponding to $\HE \lesssim 24$; however, it will be sensitive to continuum emission only from the most luminous galaxies and, so, the redshift estimation will be based primarily on the detection of emission lines \citep{2023arXiv230209372E}. The Wide Survey will be complemented by the Deep Survey, which will reach 2 magnitudes deeper in flux over an area of $50 \, \text{deg}^2$ split over three separate fields. In the Deep Survey blue grism ($[926, 1366]\,\unit{nm}$) observations will complement those with the standard red grism ($[1206, 1892]\,\unit{nm}$). Both the grisms have a dispersion of $13 \, \AA / \text{pixel}$. With greater sensitivity and an extended wavelength range, the Deep Survey will be used to construct a reference galaxy sample with secure spectroscopic redshift measurements, to characterise the selection function and redshift error distribution of the Wide Survey.

The design of the \Euclid spectroscopic survey poses a particular challenge for sample selection: bright emission-line galaxies for which the redshift can be measured make up a small fraction of all photometrically detected sources and this sample is not known beforehand. We can illustrate our expectations of the \Euclid spectroscopic sample using the Flagship2 mock galaxy catalogue, which was calibrated against the H$\alpha$ luminosity function model 3 of \citet{Pozzetti2016}. The mock catalogue contains approximately $\num{2e5} \,  \text{galaxies} / \text{deg}^2$ to the magnitude limit $\HE<24$. Out of this sample, only $2\%$ are in the redshift range $0.9<z<1.8$ and have H$\alpha$ emission-line flux greater than \num{2e-16}\,\unit{erg.s^{-1}.cm^{-2}}. The majority of the photometrically detected sources with $\HE<24$ will leave no signal on the spectrograph, being either too faint in continuum emission, or not having a detectable emission line in the wavelength range of the red grism. When targeting galaxies at the low signal-to-noise limit, spurious noise features can be mistaken for emission lines leading to wrong redshift measurements. Current end-to-end tests of the data reduction pipeline suggest that the spurious detection rate is even higher than the naive prediction based on Gaussian noise statistics due to artefacts from spectral contamination. If not appropriately treated, such wrong redshifts in the galaxy catalogue degrade the cosmological constraints derived from the two-point correlation function or power spectrum galaxy clustering statistics \citep{Addison2019}.

In principle, when selecting the sample for analysis all available information should be used to minimise the fraction of spurious measurements, while at the same time, maximising the number density of the sample, or another figure of merit.
However, the benefits from including additional constraints in the sample selection criteria must be carefully weighed against potential systematic biases. In the case of \Euclid, including additional information from ground-based photometry modifies the selection function of the survey and could couple the sample with unwanted systematic effects that arise from observations made through the Earth's atmosphere \citep[see, e.g.,][for a quantitative discussion of the impact of angular systematics on the measured clustering]{Ross2011}. The trade off of adding ground-based information will clearly also depend on the scientific analysis being considered. With slitless spectroscopy, since every galaxy in the field is in any case observed, we shall have the important advantage of being able to test a posteriori the impact of any chosen selection on the measured clustering, and evaluate the robustness of the results.

Our aim with this work is to investigate photometric classification criteria that are sensitive to both redshift and emission-line flux, in order to identify the sources that are likely to give successful spectroscopic redshift measurements in the Wide Survey. This strategy is similar to the methods used in ground-based spectroscopic surveys that make use of magnitude and colour selections to build the target sample for spectroscopy. For example, colour selections were applied to build the Sloan Digital Sky Survey Luminous Red Galaxy sample \citep{2001AJ....122.2267E} and the VIMOS Public Extragalactic Redshift Survey \citep[VIPERS;][]{2014A&A...566A.108G}. A sample of emission-line galaxies was targeted by the Extended Baryon Oscillation Spectroscopic Survey (eBOSS) using a colour selection \citep{Comparat2016}, and a similar approach was adopted for the emission-line galaxy sample targeted by the Dark Energy Spectroscopic Instrument \citep[DESI;][]{Raichoor2023}.

As a generalisation of the conventional colour cuts that are made in a two-dimensional colour-colour plane, we apply machine learning-based classification algorithms. These algorithms are well suited to optimising classification tasks in a high-dimensional parameter space. Thus, we expect them to outperform simple selection rules.

An option that is immediately available for such a use are photometric redshifts. \Euclid will construct an unprecedented photometric redshift catalogue from the combination of ground-based and \Euclid photometric bands. However, as we will discuss, photometric redshifts alone do not solve the problem. Even if photometric redshifts allow us to select a sample of galaxies at the target redshift range, additional criteria on galaxy physical properties, such as the star-formation rate, will still be needed to identify the population with bright emission lines (see Sect. \ref{sec:res-comparison}).

A schematic representation of the \Euclid spectroscopic sample selection pipeline is shown in Fig.~\ref{fig:flowchart}. A redshift measurement will be performed for all sources detected in photometry, and will be accompanied by an assessment of its confidence level, as well as the measurements of spectral features including emission-line fluxes. Sources that do not have a significant detection in spectroscopy should be assigned a low measurement confidence.
Additionally, \Euclid will produce photometric catalogues based on the $\IE$, $\YE$, $\JE$, and $\HE$-band images, which will be augmented with ground-based measurements ($u$, $g$, $r$, $i$, $z$) needed particularly for photometric redshift estimation \citep{2021ApJS..256....9S}.

The photometric classification that we discuss enters as a second input to spectroscopic sample selection. The classifier can be trained on the Deep Field catalogues, which is expected to give robust redshift measurements for the emission-line target galaxies in the Wide Survey.
The classifier will be applied to the photometric data of the Euclid Wide Survey, and its results combined with the spectroscopic measurements to build the final selected sample. This can be characterised in terms of its `redshift purity' and `sample completeness'. Any photometric criteria will necessarily reduce the number density of the sample; however, if emission-line galaxy targets can be identified from the photometry, this will increase the fraction of correctly-measured redshifts and improve the purity.

We use the terms sample completeness and redshift purity to characterise the quality of the \Euclid spectroscopic samples. We define completeness with respect to the H$\alpha$ emission-line galaxy sample that exists in the Universe, which we call the true targets.\footnote{This definition differs from that typically used in ground-based multi-object spectroscopic surveys that define completeness with respect to a known target sample constructed from photometric catalogues. Since the detection in \Euclid spectroscopy will depend primarily on the signal-to-noise ratio of the emission lines, the sample with spectroscopic redshifts will not be representative of a simple photometric selection.}
These are defined by a set of intrinsic properties, including angular position, redshift, size, and flux, that do not depend on the measurement process. Once the observations are made, we construct the sample catalogue which contains the set of measured properties, signal-to-noise estimates, and quality flags for the detected sources. The completeness tells us the fraction of the true targets that have a correct redshift measurement and make it into the sample for analysis,
\begin{equation}\label{eq:completeness}
C=\frac{N_{\text{True\,Targets\,\&\,Sample\,\&\,Correct-} z}}{N_{\rm True\,Targets}} \, .
\end{equation}
On the other hand, the redshift purity tells us the fraction of the sample that has a correct redshift measurement,
\begin{equation}\label{eq:purity}
    P=\frac{N_{\text{Sample\,\&\,Correct-} z}}{N_{\rm Sample}} \, .
\end{equation}
The redshift purity only makes reference to the sample selected for analysis and does not depend on other intrinsic properties of the galaxies besides redshift.\footnote{We do not consider the sample purity, which can include other criteria such as flux, since our main objective is to select galaxies with good redshift measurements for the galaxy clustering analysis.}

In this paper, we focus on the photometric classification, which is one step of the selection process illustrated in Fig.~\ref{fig:flowchart}.
We consider the potential gain from the photometric classification in terms of its precision and recall (defined in Sect. \ref{sec:comparison_metrics}), which will impact the final purity and completeness of the spectroscopic redshift sample. The photometric selection reduces the size of the sample in the numerator of completeness (Eq.~\ref{eq:completeness}) and thus leads to a lower value of completeness. However, it acts on both the numerator and denominator of purity (Eq.~\ref{eq:purity}), and so is a way to potentially boost the purity.
The propagation of the photometric classification to the spectroscopic sample selection and the computation of purity and sample completeness requires full end-to-end simulations of the \Euclid reduction pipeline. In Sect. \ref{sec:punco-discussion}, we will present results from preliminary simulations based on the \Euclid spectroscopic pipeline, leaving a more detailed investigation to follow-up work.

The paper is organised as follows. In Sect. \ref{sec:algorithms} we present the different algorithms we tested, and introduce the metrics we used to quantify the classifier performance. In Sect. \ref{sec:data} we discuss the mock catalogues, the noise model we apply to the photometry, and give the target definition. The results of the different analyses are presented in Sect. \ref{sec:results} and discussed in Sect. \ref{sec:res-comparison}. In Sect. \ref{sec:punco-discussion} we discuss how the photometric selection affects the spectroscopic sample. We conclude in Sect. \ref{sec:conclusions}.

\section{Classification algorithms} \label{sec:algorithms}

A classifier is an algorithm that outputs the probability of an object of being an element of a given class, or group. For the purpose of this work, which is to identify target galaxies from their photometric properties, we use a binary classifier. In this case, the algorithm simply outputs the probability $p$ of the object being a target, and $1 - p$ the probability of it being a non-target. A galaxy enters the target sample if $p > p_{\text{thresh}}$, where $p_{\text{thresh}}$ is a threshold probability value. How the threshold is chosen is discussed in Sect. \ref{sec:comparison_metrics}.

In this work we tested six different machine learning classifiers. The first three are self-organising maps (SOMs), dense neural networks (NNs) and support vector machine classifiers (SVCs). The other three are voting classifiers based on decision trees: the random forest (RF), the adaptive boosting classifier, or AdaBoost (ADA), and the extremely randomised tree classifier, or extra-tree classifier (ETC). These specific algorithms were chosen for our tests as they are known to perform well in classification tasks and are able to identify non-linear boundaries between classes.

\subsection{Classification metrics}\label{sec:comparison_metrics}

\begin{figure}
    \centering
    \includegraphics[width=.5\textwidth]{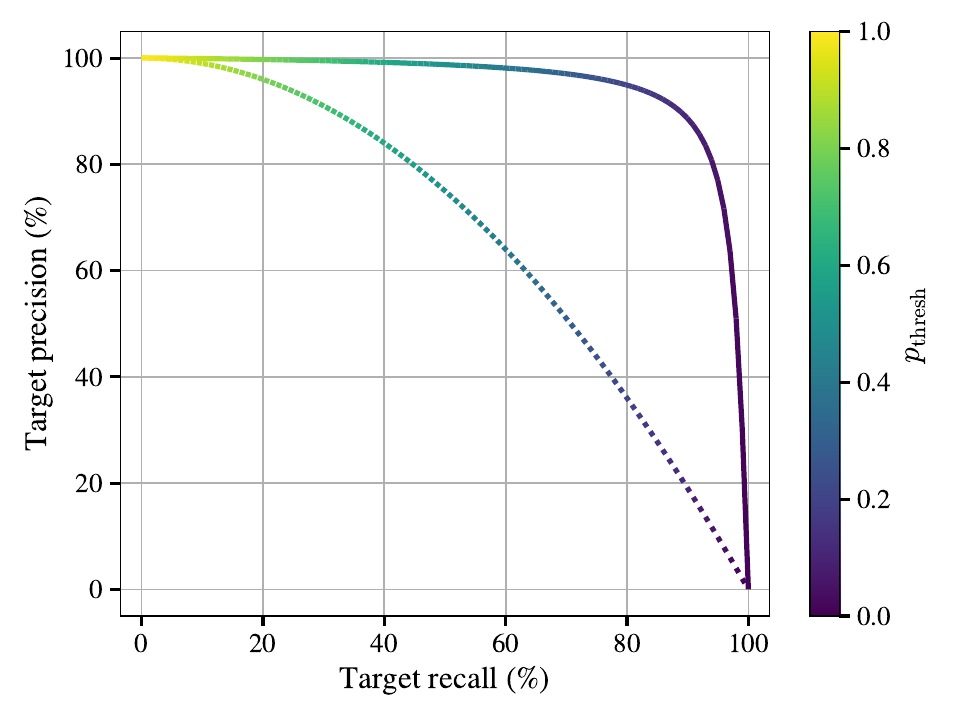}
    \caption{Relationship between precision and recall of a classifier. The lines are colour-coded as a function of the classification probability threshold. The solid and dotted lines show the behaviour of two classifiers for illustration. The classifier represented by the solid line performs better than the dotted line since it gives higher precision and recall.}

    \label{fig:pr_trhesh}
\end{figure}

To compare the results from different classifiers, we adopt three metrics defined from their `confusion matrix'. The elements of the confusion matrix of a binary classifier are the counts of true positives ($N_{\rm TP}$), true negatives ($N_{\rm TN}$), false positives ($N_{\rm FP}$), and false negatives ($N_{\rm FN}$). Our chosen metrics are the `precision', `recall', and `false positive rate' (FPR), defined respectively as
\begin{equation}
    {\rm precision}  = \frac{N_{\rm TP}}{N_{\rm TP} + N_{\rm FP}} \, ,
    \label{eq:precision}
\end{equation}
\begin{equation}
    {\rm recall} = \frac{N_{\rm TP}}{N_{\rm TP} + N_{\rm FN}} \, ,
    \label{eq:recall}
\end{equation}
\begin{equation}
    {\rm FPR} = \frac{N_{\rm FP}}{N_{\rm FP} + N_{\rm TN}} \, .
    \label{eq:FPR}
\end{equation}
The precision is the fraction of the selected sample that are true targets, i.e., it quantifies the level of contamination due to wrongly classified sources. The recall, also known as `true positive rate', is the fraction of true targets that are identified correctly (as $N_{\rm TP}+N_{\rm FN}$ corresponds to the total number of targets). The false positive rate, or `fall-out', is the fraction of non-targets that are mislabelled as targets and enter the selected sample as interlopers. The complement of the false positive rate is the `true negative rate',
\begin{equation}
    \text{TNR} = \frac{N_{\rm TN}}{N_{\rm FP} + N_{\rm TN}} = 1 - \text{FPR} \, ,
    \label{eq:TNR}
\end{equation}
which characterises the fraction of non-targets that are correctly removed from the sample.

These metrics change as functions of the probability threshold chosen for the classifier, i.e., the probability value $p_{\rm thresh}$ above which an object is classified as a target. This is a hyper-parameter of the model, which we set to maximise a chosen metric. In a binary classification, a training set is said to be `balanced' when it is evenly split between targets and non-targets, and $p_{\rm thresh} \sim 0.5$. When the training set contains a much larger number of targets than non-targets, or vice versa, it is called `unbalanced', and we refer to this case as an `unbalanced classification'. In general, in unbalanced classifications the optimal probability threshold is very different from $0.5$. Precision and recall can be computed as a function of $p_{\rm thresh}$ and plotted against each other, as shown in the example of Fig.~\ref{fig:pr_trhesh}. Such a plot is very informative for the photometric selection task that is the scope of our work. In Fig.~\ref{fig:pr_trhesh} we present two possible behaviours of this curve. The solid line is an almost ideal classifier that has high precision also when the recall is high, while the dotted curve corresponds to a classifier with worse performance. Since the photometric criteria make up only one step of the spectroscopic sample selection process (see Fig.~\ref{fig:flowchart}), we want to keep the recall of the photometric classification as high as possible. In other words, we want to get a resulting sample as complete as possible, discarding the minimum number of true targets. Thus, we choose a specific value for the recall and, from this relation, derive the corresponding precision yielded by the algorithm. We use the precision at $95\%$ recall as our benchmark value. A similar plot can be produced in terms of redshift purity and sample completeness. The shape of this curve will depend on the chosen probability threshold, and consequently the recall, of the photometric classification. In Sect.~\ref{sec:punco-discussion} we justify the choice of the $95\%$ recall value and present results for the redshift purity and sample completeness.

Finally, we use the false positive rate as the main metric to compare algorithms trained with different input features (see Sect.~\ref{sec:res-comparison}). The false positive rate helps to visualise the fraction of misidentified objects in terms of redshift or emission-line flux and shows the source of the contaminants.

\subsection{Self-organising map}

Self-organising maps \citep{kohonen-self-organized-formation-1982,SOM_1990} use unsupervised learning to project a high dimensional feature space onto a lower dimensional one, usually a two-dimensional space, as the name map suggests. We build a $55 \times 55$ map trained for 60 epochs, where an epoch corresponds to an iteration of the algorithm during which the entire training set is processed. To train the self-organising map, in addition to the photometric features used as inputs for all the other methods, we also add the target label (see Sect.~\ref{sec:data}). Then, when projecting new data onto the self-organising map the target labels are removed. These steps make the implementation of the self-organising map presented here more similar to a supervised learning algorithm. 
We also introduce a weight, $w_{\rm SOM}$, of the photometric features, which enables us to control the importance of the label in the training.  
This is a hyper-parameter of the self-organising map model. 
Finally, the probability of an object of being a target is defined by the target fraction in the cell it has been projected onto. The self-organising maps were implemented using \texttt{SOMPY} \citep{moosavi2014sompy}.

\subsection{Neural network}

\begin{figure*}
    \centering
    \includegraphics[width=1\textwidth]{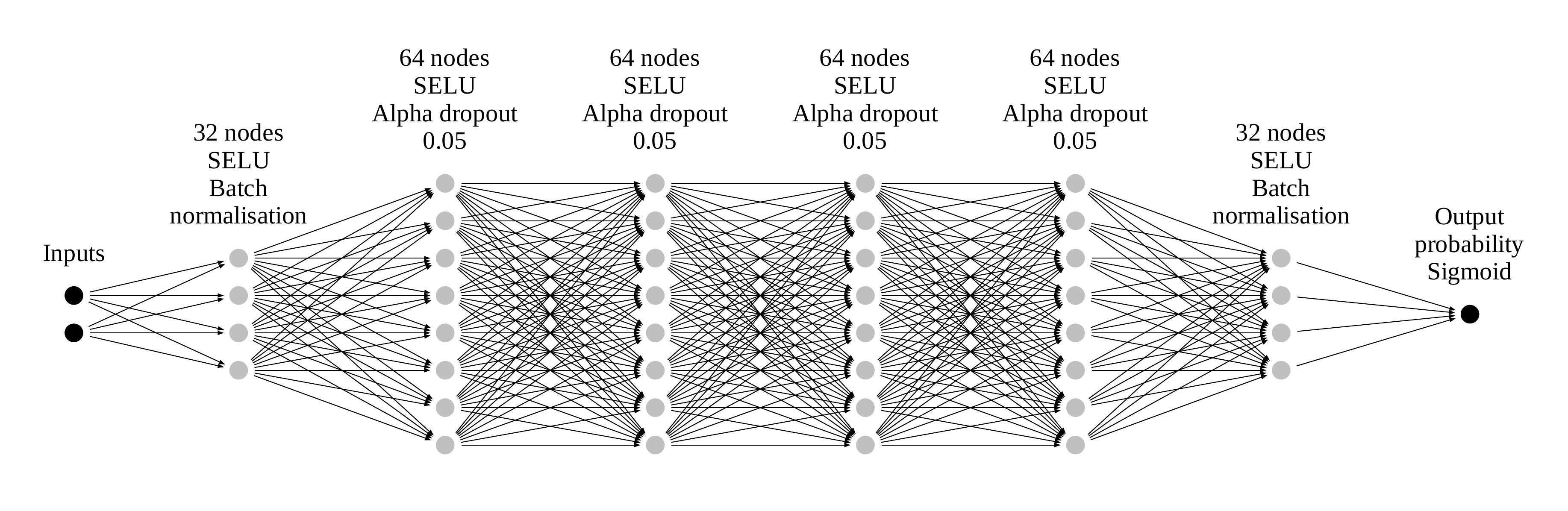}
    \caption{A schematic representation of the neural network architecture used for classification. Values pass from the input to the output along the connected edges; each node represents a linear combination of the inputs and the application of a non-linear activation function. The value at the output represents the binary classification probability between 0 and 1. The number of input neurons varies for the different configurations (4 for \Euclid-only, and 8 after adding ground-based photometry, see Sect.~\ref{sec:results}). For visualisation, the number of neurons in each hidden layer has been divided by 4.}
    \label{fig:NN-scheme}
\end{figure*}

Neural networks are by far the most popular supervised learning algorithms. 
They can be described as a sequence of layers; when the inputs are processed by a layer they first undergo a linear transformation and then a nonlinear function is applied to them. During the learning process the neural network updates the coefficients, usually called weights, of the linear transformation of each layer in order to fit the target function $y=f(x)$ that relates the inputs $x$, to the labels $y$. This structure enables neural networks to potentially fit any function of the input features \citep{LeCun15}. 

Our neural network architecture was optimised for the problem at hand. Figure~\ref{fig:NN-scheme} shows a schematic representation of the neural network. The input layer is followed by a first block that consists of a dense layer with 32 neurons and a batch normalisation layer \citep{Ioffe2015}. Then, a second block which consists of a dense layer with 64 neurons and an alpha dropout layer \citep{Klambauer2017} with rate $0.05$ is repeated four times. Finally, the first block is repeated before the output layer, which consists of 1 neuron. The activation function of all the layers except for the output is a scaled exponential linear unit \citep[SELU;][]{Klambauer2017}. The last layer, as it has to output a probability, has a sigmoid activation function. Since the ratio between positive and negative examples is very low, we opted for a sigmoid focal cross entropy loss function \citep{2017focal-loss},
\begin{equation}
    \text{FL}(p) = - \alpha \, (1 - p )^{\gamma} \ln(p) \, ,
    \label{eq:focal-loss}
\end{equation}
where $\alpha$ and $\gamma$ are two hyper-parameters of the model. We use $\alpha = 0.6$ and $\gamma = 4$. We implemented the neural network in the TensorFlow2 framework \citep{tensorflow2015-whitepaper}.

\subsection{Support vector machine classifier}

Support vector classifiers \citep[][]{Boser1992} partition the feature space by applying a kernel transformation to map curved boundaries into planes and finding the maximum-margin hyperplane that separates the classes. It is important to note that for our training we weight differently the target and non-target examples. This weighting is necessary in the case of imbalanced classes. Alternatively, one could select a balanced subsample of the original training set. However, such a solution would greatly reduce the size of the training sample. Our approach uses the support vector classifier implementation of \texttt{scikit-learn} \citep{JMLR:v12:pedregosa11a}, which has an inbuilt functionality to balance the sample via weighting.

 We adopt the \texttt{scikit-learn} default kernel, which is the radial basis function kernel (RBF), 
\begin{equation}
    K(\mathbf{x}, \mathbf{x'}) = \exp{\left(-\gamma \, \| \mathbf{x} - \mathbf{x'} \|^2 \right) \, ,}
    \label{eq:rbf}
\end{equation}
where $\| \mathbf{x} - \mathbf{x'} \|^2$ is the Euclidean squared distance and $\gamma$ is the hyper-parameter that controls the dimension of the region of influence of the training point.

\subsection{Decision tree-based classifiers}

The last three classifiers are voting or ensemble classifiers. In general, a voting classifier is an algorithm that combines the output of different base classifiers through a vote, which can be weighted or not. In this work we used classifiers based on the same base algorithm, the decision tree. These classifiers differ in how they split the data set to train the trees, how they build the trees, and how they combine together their probability outputs.

A decision tree is a supervised machine learning model that approximates a function with a series of simple decision rules \citep[see ][Chap. 9]{hastie01statisticallearning}. Decision trees have the advantages that they can be easily visualised, have high explainability, and require very little data preparation; however, they can easily over-fit the training sample making their output and final structure dependent on the training set. These issues can be reduced by combining the results of different trees \citep[e.g.,][]{bauer1999empirical}.

The first of these voting classifiers are random forests. Random forests \citep{Breiman2001} are an ensemble of decision trees each one trained with a subsample of the training set. This subsample is a bootstrap sample, which means its elements are randomly selected with replacement from the complete training set. The final output of the random forest for the classification task is a majority voting between all the decision trees of the forest. Random forests very efficiently reduce the over-fitting of single decision trees. To take into account the class imbalance of the sample we weigh the two class examples by the inverse of their frequency. The weights are computed for each bootstrap subsample.

The second ensemble classifier is a discrete adaptive boosting classifier \citep{Freund1995}. Differently from the random forest, adaptive boosting classifiers can use different base classifiers. In this work, we limited the analysis to adaptive boosting classifiers based on decision trees with weighted data to balance the sample examples. Adaptive booster classifiers combine the results of subsequently trained base learners with a weighted majority vote. At each step of the training a new learner is built from the training set, which is re-weighted to reduce the importance of data that have been correctly classified in the previous steps.

Finally, the last algorithm we use is the extra-tree classifier. Extra-tree classifiers are ensemble classifiers based on decision trees \citep{geurts2006extremely}. An extra-tree classifier is composed of a group of decision trees, which are trained with bootstrap subsamples of the training set, as in random forest training. The difference between a random forest and an extra-tree classifier lies in how the decision rules of the trees are selected. In random forests, the splits of the tree nodes are deterministic and depend on the selection algorithm; in extra-tree classifiers, instead, they are randomly drawn and the final rule is chosen as the best-performing one among them. This helps in reducing even more the variance of the method. All three voting classifiers are implemented in \texttt{scikit-learn}, and the function to weigh the data to balance them is part of their built-in functionalities.

\section{Benchmark data} \label{sec:data}

\subsection{Mock galaxy catalogues}

We use two catalogues to benchmark the selection algorithms: the EL-COSMOS catalogue and the Euclid Flagship2 mock galaxy catalogue. These catalogues include broadband photometry, emission-line fluxes and morphological properties. We make use of the \Euclid photometric bands from VIS, $\IE$, and NISP, $\YE$, $\JE$, and $\HE$ \citep{Schirmer2022}, with depths listed in Table~\ref{tab:lim_flux}. Additionally, photometric data from multiple ground-based surveys will be included in \Euclid analyses to extend the wavelength coverage to the optical with $u$, $g$, $r$, $i$, and $z$ bands and obtain reliable photometric redshifts that are key for \Euclid weak lensing science. These include the Vera C. Rubin Observatory Legacy Survey of Space and Time \citep[LSST,][]{lsst}, the Dark Energy Survey \citep[DES,][]{des}, and the Ultraviolet Near Infrared Optical Northern Survey (UNIONS).\footnote{\url{https://www.skysurvey.cc/aboutus/}.} In order to benchmark the photometric selection in this work, we use the $ugriz$ filters and UNIONS survey depths, which are listed in Table~\ref{tab:lim_flux}. Hereafter, we refer to the photometry of the four \Euclid filters as \Euclid photometry, and to the photometric data from the five optical filters as ground-based photometry. The photometry does not include the effect of Milky Way extinction.

The resolution of \Euclid NISP spectroscopic observations is not sufficient to separate H$\alpha$ from its neighbouring 
$\left[\ion{N}{ii}\right]\lambda6549$ and $\left[\ion{N}{ii}\right]\lambda6584$ companions. As such, \Euclid will measure the combined flux of this triplet of emission lines, which we shall use here and indicate for brevity as
\begin{equation}
    f_{{\rm H}\alpha+\left[\ion{N}{ii}\right]} = f_{{\rm H} \alpha} + f_{\left[\ion{N}{ii}\right]\lambda 6549} + f_{\left[\ion{N}{ii}\right]\lambda 6584} \, .
    \label{eq:flagship_el}
\end{equation}
We refer to the triplet as the `H$\alpha$ complex'. 

We also investigate the benefit of adding morphological information to the target classification. The two mock galaxy catalogues we use here include morphological model parameters including disk ellipticity, bulge scale, disk scale, and bulge-to-disk ratio; however, since these properties will not be, in general, directly measured from the data, we used them to derive the observable half-light radius, $r_{\rm half}$, and axial ratio, $e$. To do this, we ran \texttt{GALSIM}  \citep{galsim} using the morphological parameters for each mock galaxy to generate a simulated image of the galaxy as it would be observed by VIS, from which we estimated the half-light radius and axial ratio. We carried out this procedure only for the Flagship2 catalogue. 

We use only galaxies in the mock catalogue, without accounting for the possibility that stars or active galactic nuclei may be misclassified in real data and enter the sample. Contamination from faint stars, in particular, can potentially reduce the purity of the galaxy sample. The severity of such contamination depends on the performance of the star-galaxy classification, which is a separate step of the \Euclid data analysis and whose impact is beyond the scope of this work.

\subsubsection{EL-COSMOS} \label{sec:el-cosmos}

The EL-COSMOS catalogue is an extension of the COSMOS 2020 photometric
catalogue \citep{cosmos2020}. The COSMOS catalogue is a multi-band data set assembled in the \textit{Hubble} Space Telescope COSMOS field over the past fifteen years \citep{2007ApJS..172....1S}. The catalogue was extended as described in \citet{saito2021} with synthetic photometry and emission-line fluxes. To assign the fluxes of the emission lines the authors combined spectral energy fits of the stellar continuum, which correlates with the intrinsic emission line fluxes, with a careful modelling of dust attenuation as a function of redshift. 
We use an update to the emission-line catalogue produced for the Euclid Consortium (Euclid Collaboration, in prep.). It contains about $\num{2e5}$ galaxies and $\num{2000}$ active galactic nuclei. This catalogue also contains stars observed in the COSMOS field, which, as explained, we do not consider. 

\subsubsection{Euclid Flagship}

The Euclid Flagship2 mock galaxy catalogue (Euclid Collaboration, in prep.) is based on the Flagship2 N-body simulation, the large reference simulations built by the Euclid Consortium. Galaxies were added to the simulation using an extended
halo occupation distribution model. The Flagship2 galaxy mock catalogue represents an improvement with respect to the previous version in terms of modelling of the galaxy properties. The catalogue contains photometric and spectroscopic information, morphological parameters, along with lensing properties. The morphological parameters are correlated with the galaxy properties to reproduce observed trends in galaxy size. For our work, we selected a subsample of ${\sim}\,\num{2e5}$ objects that contains a number of galaxies comparable to EL-COSMOS. We note that Flagship2 does not contain active galactic nuclei, while EL-COSMOS contains about \num{2000} of them.

An additional step must be taken to compute the total flux of the H$\alpha$ complex for Flagship2 mock galaxies. The catalogue gives the flux of H$\alpha$ and of the $\left[\ion{N}{ii}\right]\lambda6584$ line only.  Assuming a relative 1:3 ratio for the $\left[\ion{N}{ii}\right]$ doublet, we estimate the total flux as
\begin{equation}
    f_{\rm H\alpha+\left[\ion{N}{ii}\right]} = f_{{\rm H} \alpha} + \frac{4}{3}f_{\left[\ion{N}{ii}\right]\lambda 6584} \, .
    \label{eq:flagship_el}
\end{equation}

The emission-line fluxes in Flagship2 were calibrated against the H$\alpha$ luminosity function model 3 of \citet{Pozzetti2016}. We use the line and broadband fluxes with internal dust attenuation applied. From Flagship2 we use both \Euclid and ground-based photometric data, as well as the morphological parameters derived as discussed earlier.

The Flagship2 catalogue also provides photometric redshift estimates, hereafter photo-$z$s, obtained with state-of-the art algorithms using both \Euclid and ground-based photometry \citep{2020A&A...644A..31E}. In order to allow the computations of photo-$z$s for billions of \Euclid sources, a two-stage approach has been adopted. First, \texttt{Phosphoros}, a template-fitting code (Paltani et al. in preparation),
is used to compute the redshift probability distribution functions on a sample of galaxies selected from reference fields that benefit from very deep observations in a large number of photometric bands \citep[e.g., COSMOS;][]{cosmos2020}. The $k$-nearest neighbour photometric redshift algorithm \citep{tanaka2018photometric}
is then used to estimate the posterior distributions of redshift for sources in the Euclid Wide Survey.
This procedure was replicated in the Flagship2 mock galaxy catalogue. In this work we use the first mode of the posterior redshift distribution as the photo-$z$ estimate. We use the photo-$z$ to select galaxies within the redshift range of interest and compare the metrics with the results from the trained classifiers.

\subsection{Noise model} \label{sec:noise}

\begin{table}
\caption{Point source magnitude limits at depth $({\rm S/N})_{\rm lim} = 10$ for $ugriz$, and for $\IE$, $\YE$, $\JE$, and $\HE$.}
\centering
    \begin{tabular}{ccc} \toprule
    Band & $m_{\rm lim,10\sigma}$ \\ \midrule
    $u$ & 23.5\\
    $g$ & 24.4\\
    $r$ & 24.1\\
    $i$ & 23.5\\
    $z$ & 23.3\\ \midrule
    $\IE$ & 24.6\\
    $\YE$ & 23.0\\
    $\JE$ & 23.0\\
    $\HE$ & 23.0\\
    \bottomrule
    \end{tabular}
\vspace{1ex}

{\raggedright \small \textbf{Notes.} The $10 \sigma$ point source depth values in AB magnitude adopted for each filter.  \par}
\label{tab:lim_flux}
\end{table}
The errors on the broadband photometric measurements were simulated assuming background-limited observations \citep{Pocino2021} such that the standard deviation on the measurement is
\begin{equation}
    \sigma_f = \frac{f_{\rm lim}}{({\rm S/N})_{\rm lim}} \, ,
    \label{eq:phot_noise}
\end{equation}
where $f_{\rm lim}$ is the flux at the specified signal-to-noise limit $({\rm S/N})_{\rm lim}$.  
In Table~\ref{tab:lim_flux} we show the AB magnitude limits, $m_{\rm lim}$, corresponding to $f_{\rm lim}$ for $({\rm S/N})_{\rm lim} = 10$.\footnote{The magnitude limits for UNIONS in Table \ref{tab:lim_flux} were computed from the $5\sigma$ limits available at \url{https://www.skysurvey.cc/survey/}.} 
The observed fluxes were then extracted from a Gaussian distribution with the true galaxy flux, $f$, as mean, and variance given by $\sigma_f$.
In order to be able to reproduce the results, we constructed observed catalogues for both EL-COSMOS and Flagship2, which contain realisations of the flux errors produced following the recipe described above.

The driving idea in the application of our selection procedure to the real \Euclid data is that the training set will be constructed from the higher signal-to-noise data of the Euclid Deep Fields, which will have high completeness and purity at the depth of the Wide Survey. In order to build a training set that matches the noise properties in the Wide Survey, the photometry from the Deep Fields will have to be either measured in Wide-like stacks or degraded appropriately to match the noise level of the Wide Survey.

\subsection{Sample selection and pre-processing} \label{sec:labels-pre}

For our analysis, we selected from the EL-COSMOS and Flagship2 catalogues two sub-samples limited to $\HE < 24$ (which corresponds to a $4 \sigma$ point-source detection limit). In addition, as mentioned earlier, the resulting Flagship2 catalogue was further sparsely sampled in order to match the same number of objects of EL-COSMOS. Each catalogue was then split into three subsets, for training, validation, and testing, containing respectively $75\%$, $15\%$, and $10\%$ of the total parent catalogue. In fact, the validation set is needed only for the training of the neural network; for the other algorithms we could use $90\%$ of the total sample as the training set. However, for the sake of a fair comparison, we opted to use the same training and test sets for all methods, by discarding the validation set objects when not needed.

The galaxies we aim to select with the photometric selection have, on top of the $\HE<24$ cut, 
\begin{equation}
    \begin{cases}
        0.9 < z < 1.8 \\
        f_{\rm H\alpha+\left[\ion{N}{ii}\right]} > \num{2e-16} \, {\rm erg} \, {\rm s}^{-1} \, {\rm cm}^{-2}
    \end{cases}
    \, ,
    \label{eq:tarets}
\end{equation}
where $z$ and $f_{\rm H\alpha+\left[\ion{N}{ii}\right]}$ are the true redshift and emission line flux of the galaxies.
The objects satisfying this selection are what we call `target' galaxies. In terms of the classifier training, we assign a label 1 to the target galaxies and a label 0 to the remaining objects, hereafter non-targets. 
It should be noted that the ${\rm H}\alpha+\left[\ion{N}{ii}\right]$ flux criterion in the target definition is specified to select galaxies with bright emission lines that are likely to give successful spectroscopic redshift measurements. We will see that this target definition does not impose a sharp flux cut in the measured sample; galaxies just below the flux limit still have a high probability of being selected and of giving a correct redshift measurement. Moreover, these galaxies will also contribute to the redshift purity metric.

The percentage of galaxies entering the target sample within the full $\HE < 24$ catalogues is very low: ${\sim} \, 8\%$ for EL-COSMOS and ${\sim} \, 3\%$ for Flagship2. The difference between the two catalogues is consistent with the current uncertainty in the H$\alpha$ luminosity function at $z>1$. The low target fractions of the two catalogues make the classification task extremely unbalanced. The solutions adopted for each classifier were discussed in Sect. \ref{sec:algorithms} and span from weighting schemes to specific loss functions.

Finally, all input training parameters are pre-processed via standard scaling,
\begin{equation}
    X = \frac{x - \bar{x}}{\sigma_x} \, ,
    \label{eq:scaling}
\end{equation}
where $\bar{x}$ is the mean value of input feature $x$ over the training sample, and $\sigma_x$ its standard deviation. After this normalisation, the sample has zero mean and unit standard deviation, which makes the training of the algorithms more efficient, typically leading to better results.

\section{Results and discussion} \label{sec:results}
\begin{table}[]
\caption{Recall ($\%$) and precision ($\%$) for different $\HE$ cuts.}
    \centering
    \begin{tabular}{ccccc} \toprule
        & \multicolumn{2}{c}{EL-COSMOS} & \multicolumn{2}{c}{Flagship2} \\ \midrule
         $\HE$ cut & Recall & Precision & Recall & Precision \\ \midrule
         22.84 & 95 & 13.8 & - & - \\ \midrule
         22.06 & - & - &   95 & 8.9 \\ \midrule \midrule
         21.0 & 20.2 & 11.3 & 47.6 & 10.0 \\ \midrule
         22.0 & 63.1 & 16.3 &  93.3  &  9.1 \\ \midrule
         23.0 & 97.1 & 12.7 &  100  & 4.8  \\ \midrule
         24.0 & 100 & 7.8 &  100  & 2.6  \\  \midrule \midrule
         Colour cut & 95 & 14.3 & 95 & 9.9 \\ \bottomrule
    \end{tabular}
    \vspace{1ex}

{\raggedright \small \textbf{Notes.} Precision and recall for EL-COSMOS and Flagship2 at different $\HE$ magnitude limits. The first two rows correspond to the $\HE$ cut that gives $95\%$ recall respectively for EL-COSMOS and Flagship2. \par}
    \label{tab:Hcut95recall}
\end{table}
\begin{table*}[]
    \caption{Precision values ($\%$) at $95\%$ recall for the different classifiers.}
    \centering
    \begin{tabular}{cccccccc} \toprule
    & & SOM & NN & SVC & RF & ADA & ETC \\  \midrule
    \multirow{2}{4em}{\Euclid} & EL-COSMOS & 13.9 & 17.5 & 17.3 & 16.4 & 12.9 & 16.7 \\ \cmidrule(l){2-8}
    & Flagship2 & 12.7 & 16.0 & 18.0 & 15.5 & 10.4 & 16.9 \\ \midrule
    & Flagship2 morphology & 9.6 & 17.6 & 16.8 & 15.3 & 11.0 & 14.7 \\ \cmidrule(l){2-8}
    \Euclid + & EL-COSMOS ground & 20.7 & 34.3 & 34.3 & 31.5 & 29.1 & 28.0 \\ \cmidrule(l){2-8}
    & Flagship2 ground & 26.1 & 47.9 & 43.5 & 39.3 & 39.7 & 35.6 \\ \bottomrule
    \end{tabular}
    \vspace{1ex}

{\raggedright \small \textbf{Notes.} The two top rows give the results for training using \Euclid photometry only, while morphological data and ground-based photometry, respectively, are used in the bottom rows. The relative uncertainty on all values is ${\sim} \, 6\%$, estimated from multiple realisations of the training and test sets. \par}
    \label{tab:precision95recall}
\end{table*}

\subsection{Benchmark selections} \label{sec:colour-selection-benchmark}

\begin{figure*}
    \centering
    \includegraphics[width=.5\textwidth]{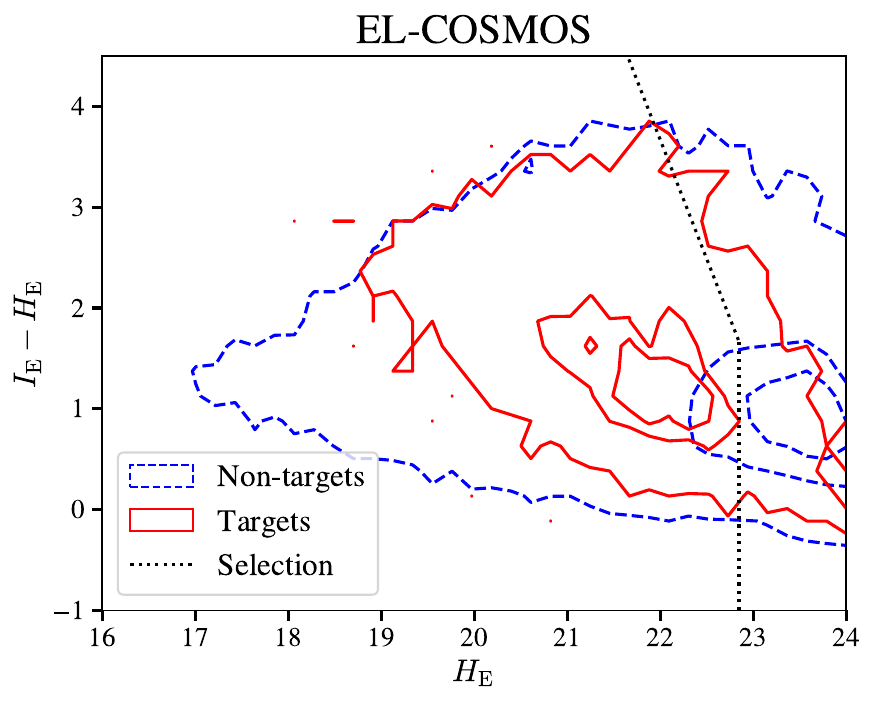}\includegraphics[width=.5\textwidth]{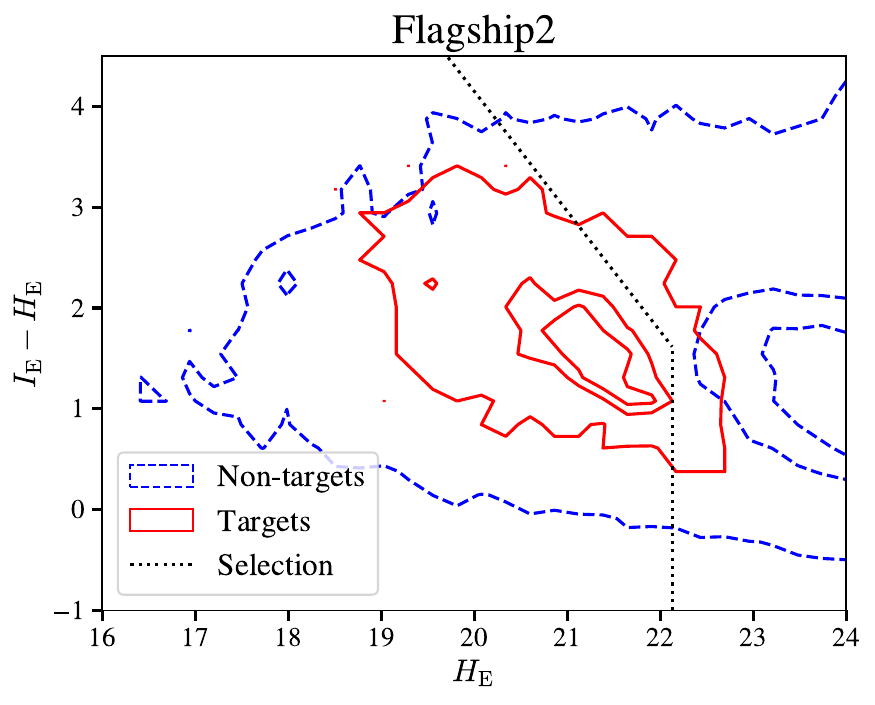}
    \caption{Optimised colour selection in the $(\IE - \HE)$ versus $\HE$ colour-magnitude plane for EL-COSMOS and Flagship2. The blue dashed lines correspond to the non-target distribution and the solid red lines to the target distribution. The contours contain $99\%$, $50\%$, and $25\%$ of the samples. The dotted black segments represent an optimised colour cut in this plane corresponding to recall ${\sim}\,95\%$.}
    \label{fig:colour-selection}
\end{figure*}

Before discussing the performance of the machine learning algorithms we present the results from simple classifiers based on magnitude and colour with \Euclid photometry. These tests provide a 
benchmark for the machine learning classifiers. We focus on the $(\IE - \HE)$ versus $\HE$ plane, which shows the largest displacement between targets and non-targets 
(see Figs.~\ref{fig:colour-selection} and \ref{fig:color-color}).
The distributions are seen to be most separated in $\HE$ magnitude.
Indeed, the use of $\HE$ is expected to be particularly suited to capture information on the H$\alpha$ flux, as it covers the $[1.5,2.0]\, \micron$ band, which encompasses the H$\alpha$ complex for $1.3 \lesssim z \lesssim 2$. In addition, the $(\IE - \HE)$ colour is sensitive to redshift, since it spans the $4000\,\AA$ break at $z>1$.

We thus begin by considering magnitude-limited samples in $\HE$. Table~\ref{tab:Hcut95recall} gives the resulting recall and precision metrics for $\HE$ cuts ranging from 21 to 24 magnitudes. For the Flagship2 catalogue, all targets have $\HE < 23$ giving  100\% recall at that limit, while for EL-COSMOS, 100\% recall is reached at $\HE < 24$.

Next, we consider a selection in the $(\IE - \HE)$ versus the $\HE$ plane.
The colour-magnitude selection reads as follows,
\begin{equation}
    (\IE - \HE) < a \, (\HE - b )  \quad \text{AND} \quad \HE < \HE^{\text{cut}} \, .
    \label{eq:colour-cut}
\end{equation}
We searched for a selection with the form of Eq.~\eqref{eq:colour-cut} that maximises the purity while giving recall ${\sim}\,95\%$, which we chose as the reference value for comparing the algorithms (see Sects.~\ref{sec:comparison_metrics} and \ref{sec:punco-discussion}). 
The best colour cut for EL-COSMOS has slope $a = - 2.36$, pivot $b = 23.60$, and $\HE^{\text{cut}} = 22.85$. For Flagship2 the slope is $a = - 1.90$, $b = 14.74$, and $\HE^{\text{cut}} = 22.13$.

Figure~\ref{fig:colour-selection} shows the targets (solid red) and non-target (dashed blue) distributions in the colour-magnitude plane of interest for EL-COSMOS (left panel) and Flagship2 (right panel). The dotted black line corresponds to the colour-magnitude cut. The two panels show the difference in the target distributions of EL-COSMOS and Flagship2. Flagship2 does not have any targets with $\HE > 23$, in contrast, EL-COSMOS targets reach the magnitude limit of the sample. For this reason we allow the $\HE$ cut to adapt to the training data. We report the precision of the optimised colour cuts in the bottom row of Table~\ref{tab:Hcut95recall}. 

From Table~\ref{tab:Hcut95recall}, we see that all selections give a higher precision for EL-COSMOS than Flagship2. This can be understood since the fraction of targets is higher in EL-COSMOS than in Flagship2. The colour cut gives a marginal improvement in purity ($0.5$ -- $1$) over the $\HE$ cut. We next show the results from machine learning classifiers, which make full use of the high-dimensional parameter space to optimise the selection.

\subsection{Using \Euclid data only}\label{sec:euclid}
\begin{figure*}
    \centering
    \includegraphics[width=0.5\textwidth]{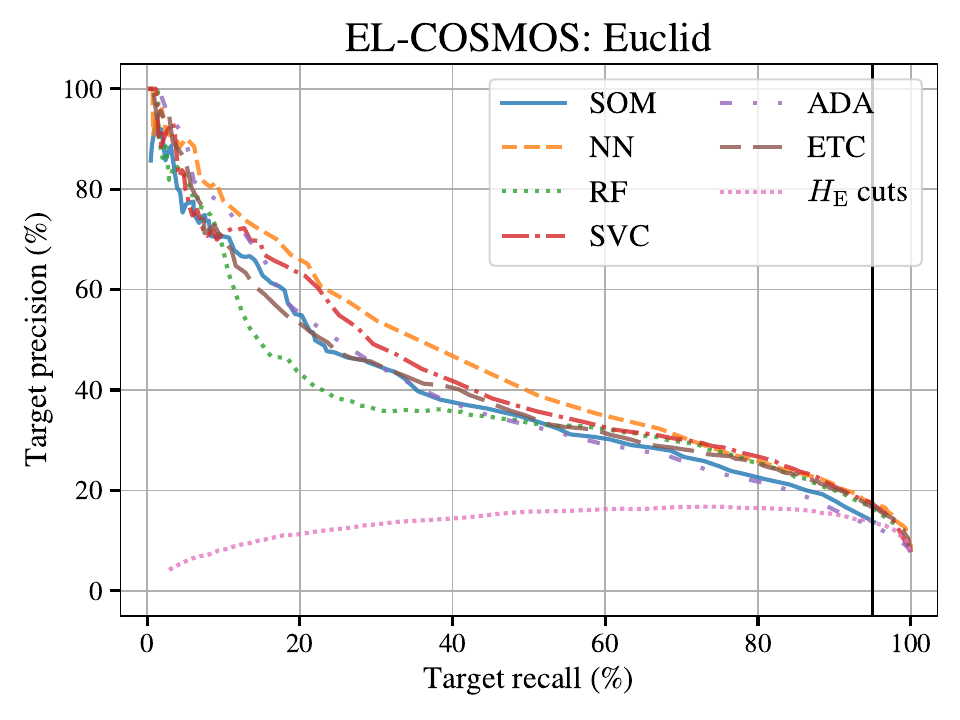}\includegraphics[width=0.5\textwidth]{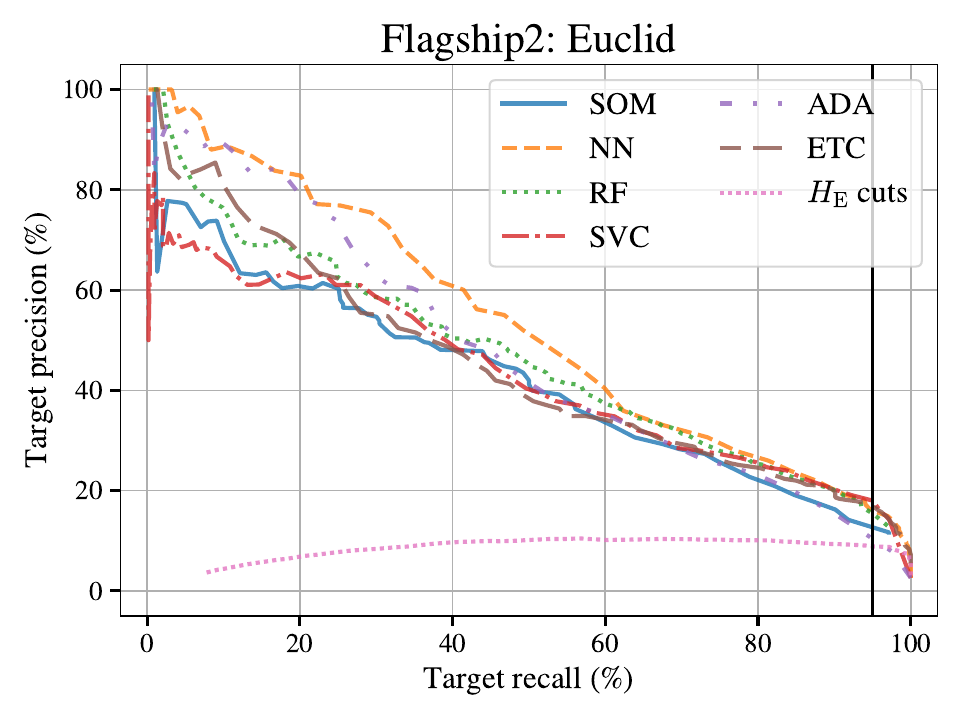}
    \caption{Precision vs. recall performance of the different classifiers, using \Euclid photometry alone for the training. The two panels correspond to the two test catalogues as indicated. The vertical solid line gives our reference recall value of $95\%$.}
    \label{fig:pr_n}
\end{figure*}

We first discuss the results obtained training the classifiers using only \Euclid photometry, comparing the two catalogues EL-COSMOS and Flagship2. The input features for each object are the same for both catalogues, namely its $\HE$ magnitude and near-infrared colours, $(\IE - \YE)$, $(\YE - \JE)$, $(\JE - \HE)$. 

Figure~\ref{fig:pr_n} shows the precision-recall curves produced by the six different classifiers using respectively EL-COSMOS (left panel) and Flagship2 (right panel). We remark that for an ideal classifier the plot would show a close-to-flat precision around unity (see Fig.~\ref{fig:pr_trhesh}), followed by a sharp drop at the highest possible recall value. To provide a reference baseline, in Fig.~\ref{fig:pr_n} we also present (dotted magenta line) the curve one obtains when simply selecting $\HE < \HE^{\text{limit}}$ magnitude-limited samples. The curve has been computed by smoothly varying $\HE^{\text{limit}}$ between $20.0$ and $24.0$ (see Table~\ref{tab:Hcut95recall}).
The vertical black line corresponds to $95\%$ recall, as reported in Table~\ref{tab:precision95recall}. 

Comparing the two panels, the first evident difference is the larger variance in performance over the whole recall range shown by the different algorithms in the case of the Flagship2 sample. Conversely, the classifiers trained with EL-COSMOS show a sharper drop in precision at small recall values.
In both cases, the magnitude-limited selection is (not unexpectedly) worse than the machine learning classifiers, but in the case of EL-COSMOS the resulting performance becomes comparable to that of the worse-performing classifiers at 95\% recall.

Overall, Fig.~\ref{fig:pr_n} and Table~\ref{tab:precision95recall} show similar performance when training with either Flagship2 or EL-COSMOS, with the former showing a larger variance at the recall threshold. Such an agreement is an encouraging indication of the robustness of the general conclusions that can be drawn from these results. In both cases, the best-performing algorithms are the neural network, the support vector classifier, and the extra-tree classifier. The random forest follows shortly behind, indicating that the bootstrap resampling used in the decision tree training is especially efficient for this task. Last comes the self-organising map, which is not optimised for this kind of task, and the adaptive boosting classifier.

\begin{figure}
    \centering
    \includegraphics[width=.5\textwidth]{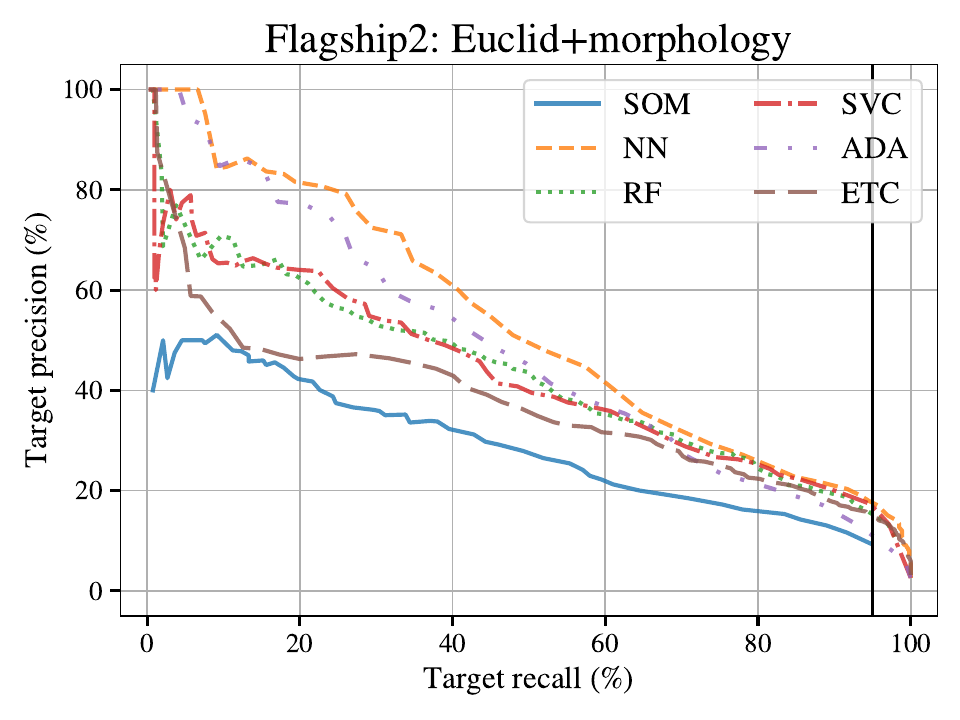}
    \caption{Same as Fig.~\ref{fig:pr_n}, but now adding morphological information in terms of half-light radius and axial ratio values.}
    \label{fig:pr_euc-morph}
\end{figure}
 
The effect of complementing \Euclid infrared photometry with morphological information described by the galaxy half-light radius and axial ratio values can be seen in Fig.~\ref{fig:pr_euc-morph}. The plot shows no large improvement and some classifiers perform worse. We also observe an even larger variance between the different classifiers, especially at low recall values. The best-performing one is still the neural network, followed by the support vector classifier, the random forest, and the extra-tree classifier. Again, the adaptive boosting classifier and the self-organising map fare poorly. A more detailed discussion is left for Sect.~\ref{sec:res-comparison}.

\subsection{Adding ground-based photometry} \label{sec:Euclid+GB}

\begin{figure*}
    \centering
   \includegraphics[width=.5\textwidth]{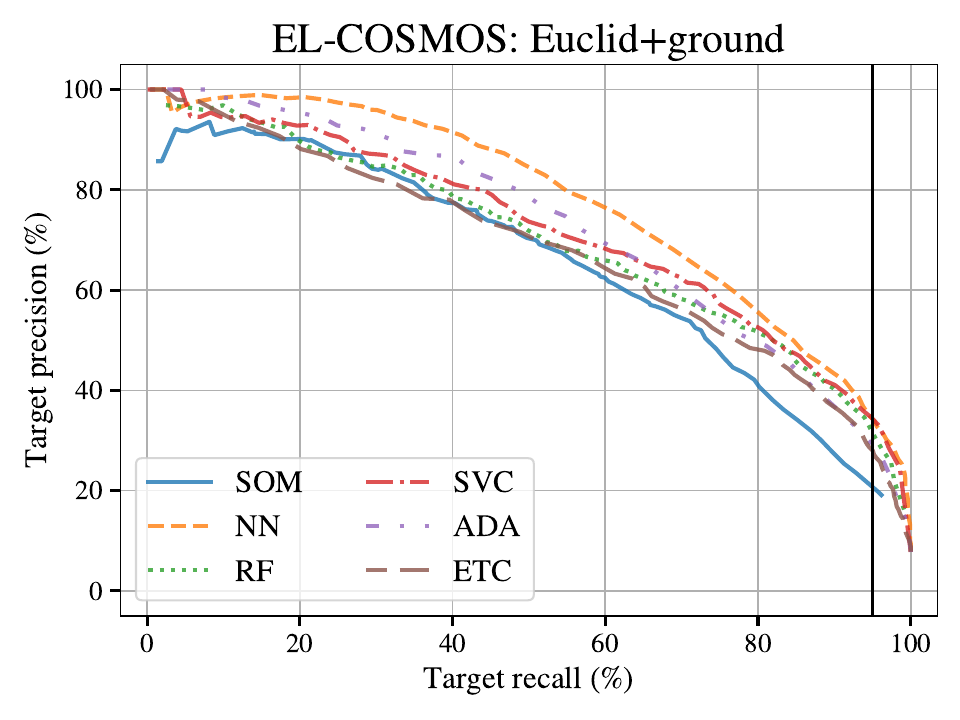}\includegraphics[width=.5\textwidth]{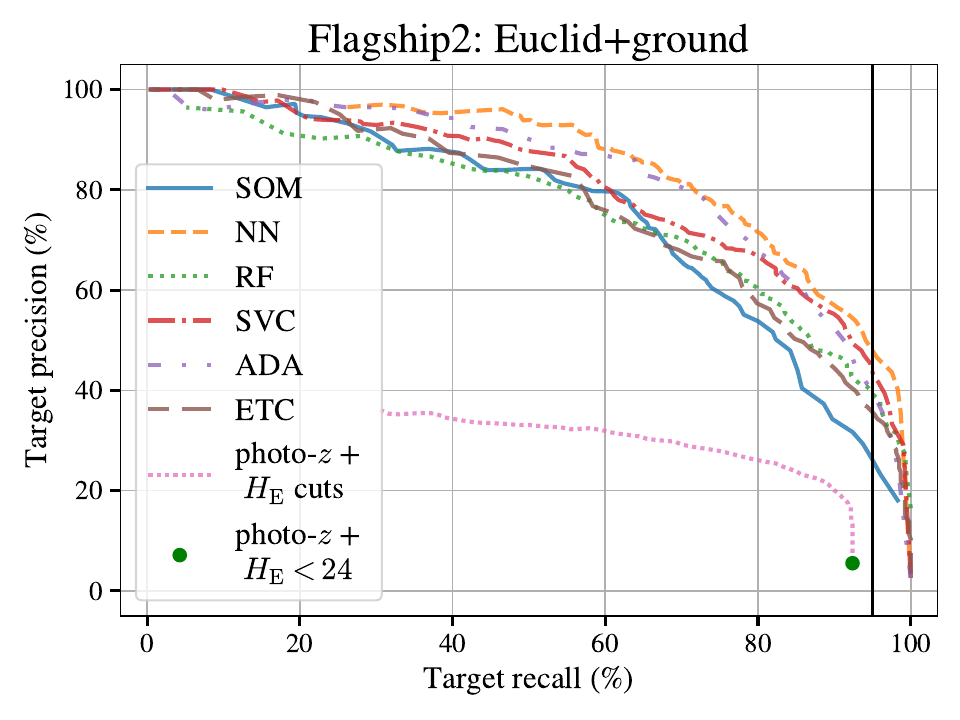}
    \caption{Precision versus recall curves for the analyses with \Euclid photometry and ground-based photometry. \emph{Left}: results for EL-COSMOS. \emph{Right}: results for Flagship2. The dotted magenta line represents the photo-$z$ selection for a range of $\HE$ magnitude limits. The green marker indicates the photo-$z$ selection with $\HE<24$.}
    \label{fig:pr_WW}
\end{figure*}

When we combine \Euclid and ground-based photometry we substitute the $\IE$ band with the five ground-based filters, $ugriz$. In this case, the input features of the classifiers are the following seven colour combinations $(u-g)$, $(g-r)$, $(r-i)$, $(i-z)$, $(z-\YE)$, $(\YE-\JE)$, and $(\JE-\HE)$. In addition, we also use $\HE$ as the last input feature.

Figure~\ref{fig:pr_WW} shows how the diagnostic plots change when combining \Euclid and ground-based photometry. 
We immediately see from Fig.~\ref{fig:pr_WW} how the ground-based data improves the overall performance, yielding curves that are much closer to the ideal shape (see Fig.~\ref{fig:pr_trhesh}). In the case of Flagship2 we also mark (green dot) the precision (${\sim}\,5.5\%$) and recall (${\sim}\,92.4\%$) values recovered when using photometric redshifts to simply isolate targets with $0.9\le z_{\rm photo} \le 1.8$, with no extra information to constrain the desired H$\alpha$ line flux. We note that the photometric redshift selection does not reach the $95\%$ recall value. We also consider photometric redshift selections with various $\HE$ magnitude limits, shown by the magenta dotted line. In Appendix~\ref{app:photz-feat} we present a preliminary test that combines the photometric and the redshift information in the training of a neural network. 

The improvement in performance appears to be larger when estimated using Flagship2 than with EL-COSMOS with a difference of ${\sim} \, 10\%$ in precision for all algorithms. The reason for this can be related to the colour distribution of the targets. In the EL-COSMOS catalogue the distribution functions of magnitudes and colours for targets shows more variance than in Flagship2 where the targets are more localised on colour space.  When \Euclid-only photometry is used, the information is not sufficient for tightly constraining the target region in the parameter space, thus producing similar results from the two catalogues. However, when ground-based photometry is added, in the Flagship2 case it becomes easier to isolate the targets. These differences may be due to the recipes used for assigning spectral energy distributions and synthetic emission lines in the two catalogues.

The relative ranking of the different classifiers derived from the two catalogues is the same. The worst performing algorithm at the recall threshold is the self-organising map, which shows a steeper drop in precision than the others (see Table~\ref{tab:precision95recall}). The remaining algorithms have precision values $>35 \%$ for Flagship2, with the neural network reaching almost $50\%$. For EL-COSMOS at the recall threshold the values of the precision are always $>25\%$, peaking at ${\sim} \, 34\%$ for both the neural network and the support vector classifiers. 

\subsection{Comparison of the results} \label{sec:res-comparison}

In this section, we focus on the results based on the Flagship2 training and discuss the results obtained with the three configurations. We will then focus on the best-performing classifier, the neural network, and discuss in more detail the three cases. We will also show a comparison with a simpler redshift-only selection based on \Euclid photometric redshifts.

Figures~\ref{fig:pr_n}, \ref{fig:pr_euc-morph}, and \ref{fig:pr_WW}, together with Table~\ref{tab:precision95recall}, provide a direct quantitative comparison of the three training configurations: the best performance is obtained by the combined \Euclid and ground-based photometry. For Flagship2, this more than doubles the precision at the recall threshold with respect to the other two configurations, a clear benefit of the extra information on lower redshift objects provided by the optical bands (see discussion in the following). The addition of morphological information through the half-light radius and ellipticity, conversely, does not introduce any significant improvement: the neural network and the adaptive boosting classifier show only a minimal gain, while all others worsen their performance. 

The half-light radius, in particular, does show a trend as a function of redshift, but this relation has a large scatter and weak correlation coefficient. It is possible that other morphological measures that we did not consider, such as the S\'{e}rsic index, will be more sensitive to galaxy type and have a greater importance for classification; however, we reserve this investigation for future work.
When fed uninformative features, the classification algorithms will tend to ignore them.
The majority of the tested classifiers have, in fact, built-in mechanisms to ignore a feature. Specifically, the neural network would reduce, during the training, the weight of the specific feature that appears to be uninformative, while the decision tree-based classifier would not introduce decision rules based on it. Similarly, the support vector classifier would only produce boundaries orthogonal to an uninformative feature. The same cannot be said about a standard self-organising map: in this case, the effect of an uninformative feature is to spread the classification targets over a larger number of cells, thus reducing the sensitivity.

\begin{figure}
    \centering
    \includegraphics[width=.5\textwidth]{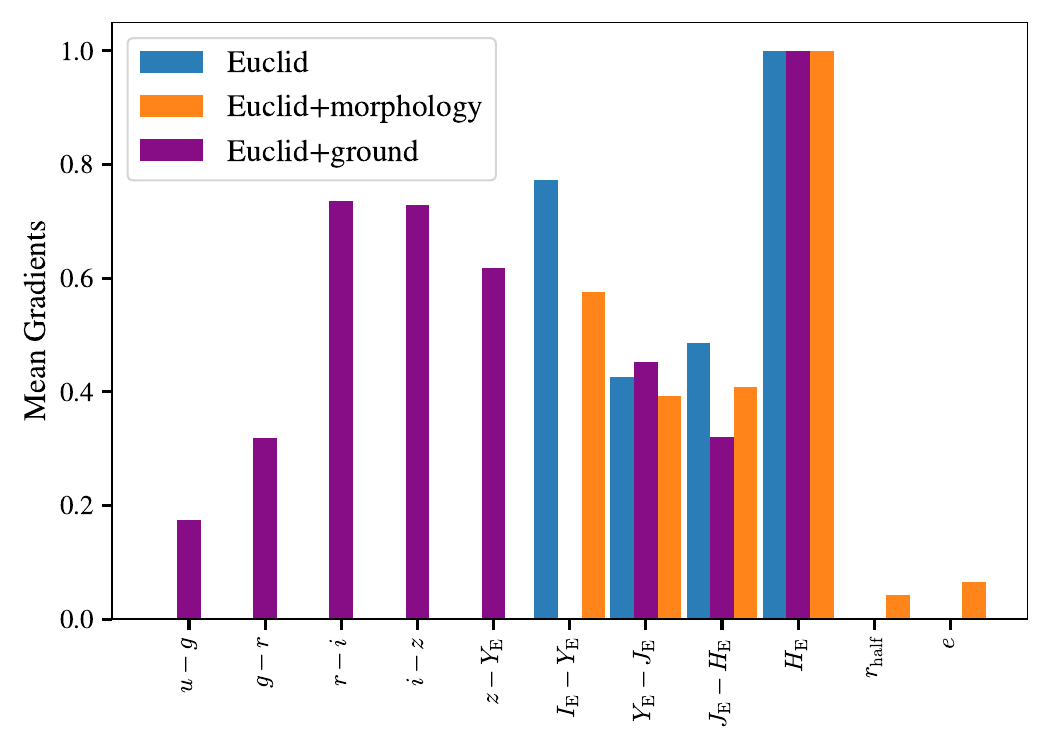}
    \caption{Mean gradients of the neural network output as a function of the input for the three training configurations. In blue, orange, and purple are respectively plotted the mean gradients of the neural networks trained with \Euclid photometry, \Euclid photometry and morphology, and \Euclid and ground-based photometry. All gradients have been normalised to that corresponding to the \Euclid $\HE$ magnitude.}
    \label{fig:gradients}
\end{figure}

In order to understand which features are most relevant for classification, which is known as the `saliency' in the machine learning literature, in Fig.~\ref{fig:gradients} we show the mean gradients of the network outputs with respect to the input features.
We see that the most important feature turns out to be the $\HE$ magnitude, followed by the ground-based colours. The dependence on the optical colours and in particular on $(\IE - \YE)$ in the \Euclid photometry configuration has two main reasons. First, the optical bands retain low redshift information (see following discussion); second, the correlation between the $\IE$ and $f_{\rm H\alpha+\left[\ion{N}{ii}\right]}$ is even stronger than the correlation of the emission line flux and $\HE$. The network uses $(\IE - \YE)$ to extract $\IE$ from the pivot magnitude $\HE$ and infer this correlation. Lastly, as expected, the morphological parameters are the least important inputs for the neural network. 

\begin{figure}
    \centering
    \includegraphics[width=.5\textwidth]{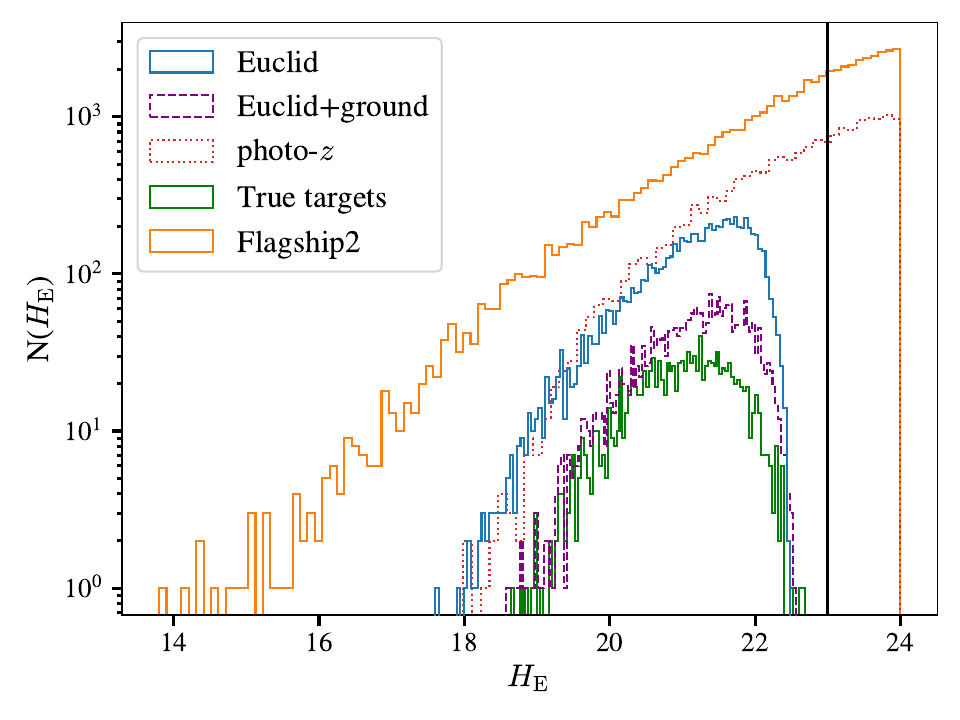}
    \caption{$\HE$-band number counts for samples built from the Flagship2 catalogue. The samples selected with the neural network classifier, using \Euclid photometry only or combined with ground-based photometry are shown, respectively, by the blue-solid and magenta-dashed histograms. As indicated by the legend, the red-dotted histogram corresponds to a sample selected in redshift only, using \Euclid photometric redshifts. The counts for the full  Flagship2 catalogue and the true target sample are also shown for reference, by the orange and green histograms. Note how the distribution of the true targets (green histogram) dies off at magnitudes fainter than $\HE\simeq 22.5$. The targets in the EL-COSMOS catalogue extend to fainter flux.}
    \label{fig:pH}
\end{figure}

Having identified the $\HE$ magnitude as the most informative feature, we can gain additional intuition about the classifiers by comparing the number counts $N(\HE)$ of the true targets to those of the samples recovered by the neural network. These are shown in Fig.~\ref{fig:pH}. The green histogram gives the number counts for the true targets, i.e., the reference distribution we are trying to reproduce with the classifier. Notably, the counts go to zero for $\HE>22.5$, hence there are no target galaxies fainter than this magnitude. This explains the rapid gain in precision one obtains by simply cutting the full sample (here shown by the orange histogram) at brighter and brighter values of $\HE$ (see Table~\ref{tab:Hcut95recall}). Looking at the other histograms, we see that the application of the neural network effectively cuts the distribution down to the correct $\HE$. When using only \Euclid bands (blue histogram), this leaves an excess of sources, which are either outside the redshift range or below the chosen H$\alpha+\left[\ion{N}{ii}\right]$ flux limit, which are significantly reduced by adding the ground-based information (magenta dashed histogram). Note also how a selection over the target redshift range $[0.9,1.8]$ using photometric redshifts clearly does not effectively cut on the $\HE$ magnitude, leaving a large population of faint objects. Nevertheless, we remind the reader that this discussion is specific to Flagship2. In the case of EL-COSMOS, also galaxies fainter than $\HE \simeq 22.5$ are part of the target sample (see Fig.~\ref{fig:colour-selection} and Table~\ref{tab:Hcut95recall}).

\begin{figure}
    \centering
    \includegraphics[width=.5\textwidth]{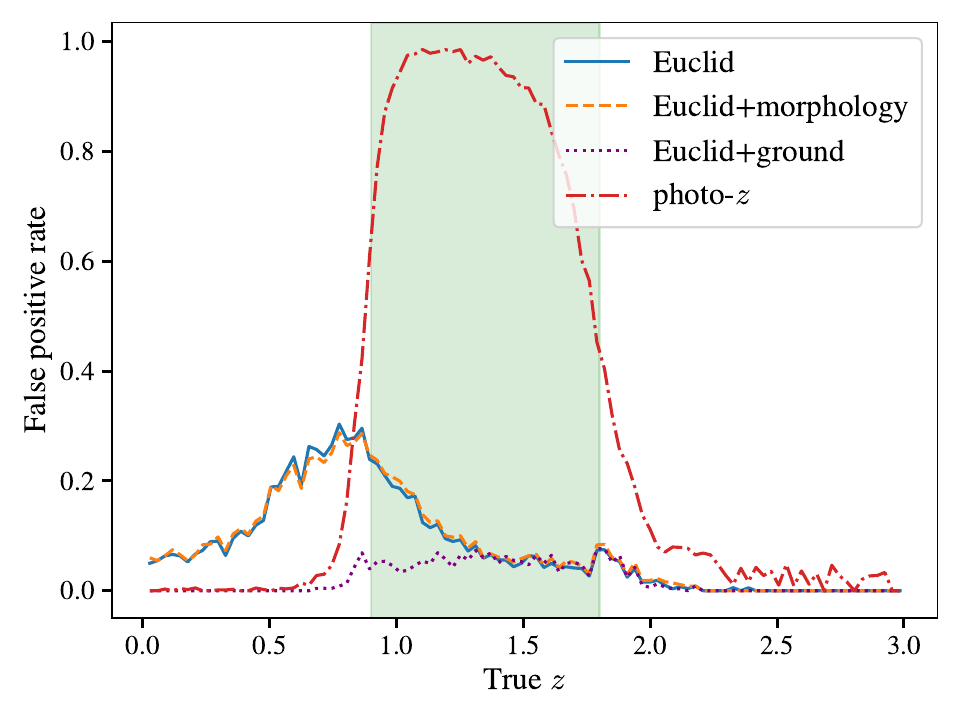}
    \includegraphics[width=.5\textwidth]{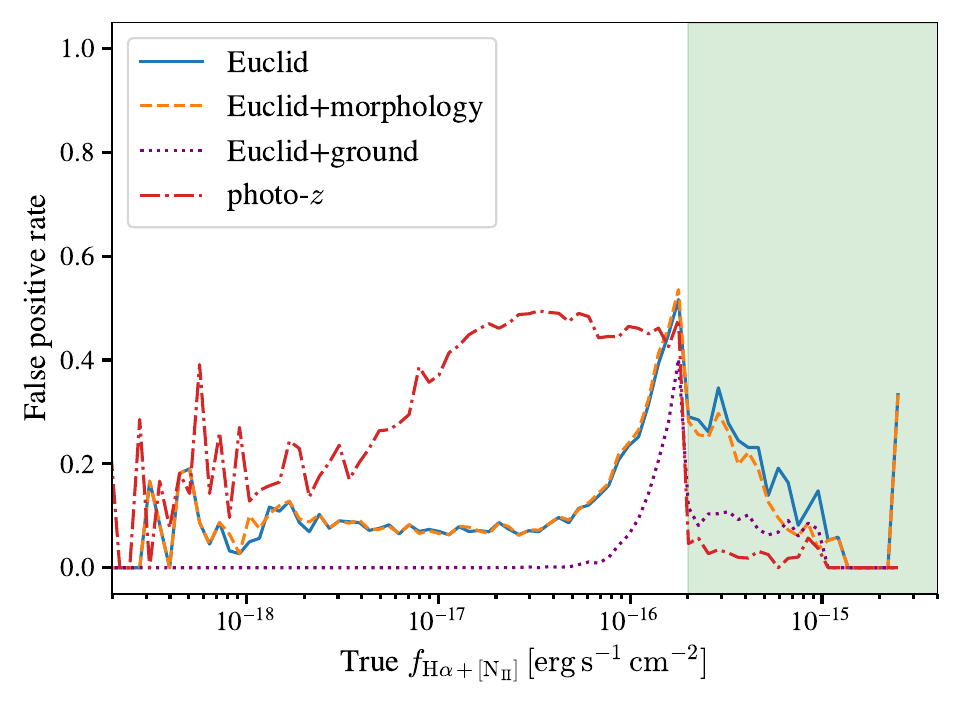}
    \caption{False positive rate as function of redshift and $f_{\rm H\alpha+\left[\ion{N}{ii}\right]}$. The plot allows us to identify the origin of non-targets that enter the selected samples. The solid blue, dashed orange, and dotted purple curves, respectively, correspond to the neural networks trained with \Euclid photometry, \Euclid plus morphological data, and \Euclid plus ground-based photometry. The dash-dotted red line is the false positive rate of the photo-$z$ selection. \emph{Top}: False positive rate as a function of $z$. The green shaded area marks the target redshift range. \emph{Bottom}: False positive rate as a function of $f_{\rm H\alpha+\left[\ion{N}{ii}\right]}$. The green shaded area corresponds to the H$\alpha$ limiting flux. There is a peak in the false positive rate just below the flux limit used to define the target sample, although we note that these galaxies can still give correct redshift measurements.
    }
    \label{fig:FPR}
\end{figure}

We can use the false positive rate (see Sect.~\ref{sec:comparison_metrics}) to interpret the origin of misclassified galaxies as a function of redshift and emission-line flux. In the top panel of Fig.~\ref{fig:FPR} this quantity is plotted as a function of redshift. The sample produced using \Euclid photometry alone shows an excess of false positives at $z<1$. This explicitly shows the inability with only the \Euclid bands to properly exclude low-redshift galaxies, as well as some with flux below the flux limit. The addition of ground-based photometry effectively cures this, removing all galaxies at $z<0.9$, leaving only a fraction of misidentified objects fainter than $\num{2e-16} \, {\rm erg} \, {\rm s}^{-1} \, {\rm cm}^{-2}$ inside the target redshift range. It is interesting to note that the \Euclid-only and the \Euclid plus ground curves become indistinguishable at $z \gtrsim 1.4$. This is consistent with the redshift at which the $4000 \, \AA$ break enters the $\YE$ band (at $9600 \, \AA$) and indicates that in this range the combination of $\IE$ and $\YE$, $\JE$, and $\HE$ provides, in general, sufficient spectral leverage to break degeneracies to both capture the correct redshift and identify emission-line targets. As also shown, a photometric redshift selection is effective at removing low-redshift galaxies, but keeps in the sample all the low-flux galaxies (as is expected, since we are selecting on redshift alone).

In Fig.~\ref{fig:FPR} bottom panel, instead, we plot the false positive rate as a function of $f_{\rm H\alpha+\left[\ion{N}{ii}\right]}$. In this case, the peak and discontinuity evident at the flux limit, $\num{2e-16} \, {\rm erg} \, {\rm s}^{-1} \, {\rm cm}^{-2}$, is due to sources just below the flux limit, which enter the sample as false positives. Above the flux limit, instead, false positives arise from galaxies that are outside the redshift range. For this reason, the photo-$z$ selection gives the lowest false positive rate, followed by the \Euclid and ground-based classification. This does not tell the full story, however. The photo-$z$ selection includes a number of false positives entering the sample at low fluxes, which are the cause of the very low precision shown by this selection. The classifier trained with ground-based photometry provides the best solution by balancing the two conditions of removing objects below the line flux limit and outside the redshift range.

Complementarily, it is also interesting to look at the true negative rate (Eq.~\ref{eq:TNR}) of the whole selected sample, which gives an insight into the fraction of non-targets removed from the sample. When we select galaxies using \Euclid photometry only, the true negative rate is $87\%$; the combination with ground-based data increases this metric up to $97\%$. Conversely, the true negative rate of the photo-$z$ selection is $59\%$. Again, the better performance of the classifiers in comparison to the photo-$z$ selection reflects the fact that the latter does not make a selection in the emission-line limiting flux.

Finally, the machine learning algorithms identify regions in the full colour-magnitude space with a higher density of targets. In the case of the classifier trained on \Euclid photometry, this is a four-dimensional space. In Appendix \ref{sec:probmaps} we present slices through the four-dimensional probability maps constructed from each classifier, showing how the selection depends on colour. It is interesting to visualise the boundaries constructed by each classifier. There is no visible separation between target and non-target galaxies in the colour planes and the classification algorithms define complex boundaries in the four-dimensional space. The support vector classifier and the neural network produce particularly smooth boundaries, while the self-organising map and tree-based classifiers do not. The irregular boundary is an indication that the classifier is over-fitting the training set and will not generalise well. In addition, we verified that the $5\%$ of the targets that we lose by imposing the $95\%$ recall value are uniformly distributed in colour and are not part of any particular object class. We note that the lost targets are mainly faint objects.

\section{Purity and completeness} \label{sec:punco-discussion}

The final purity of the spectroscopic sample will depend on the combination of the photometric information with the selection criteria applied to the spectroscopic measurements, as described by the flow diagram of Fig.~\ref{fig:flowchart}. To provide a concrete, yet preliminary, example, we would like to quantify here the improvement in the final redshift purity and sample completeness produced by our photometric selection process. This work is based on a set of simulated spectra that were processed by the \Euclid spectroscopic measurement pipeline (the SPE processing function). Although the simulated data were not yet fully realistic, they are nevertheless very useful for understanding how a machine learning-based photometric classification can aid in the sample selection. Also, the simulated spectra were built from the EL-COSMOS sample described in Sect.~\ref{sec:el-cosmos}, which helps in making this test self-consistent. Two-dimensional spectral images were generated using the \texttt{FastSpec} code based on the spectral energy distribution and morphological parameters of the galaxies. These images were convolved with the NISP instrumental point spread function and realistic noise was added according to the detector model. Multiple exposures were simulated for each source and stacked with one to four exposures. One-dimensional spectra were extracted from the images and input to the \Euclid spectroscopic measurement processing function to measure the redshift and spectral features.

\begin{figure}
    \centering
    \includegraphics[width=.5\textwidth]{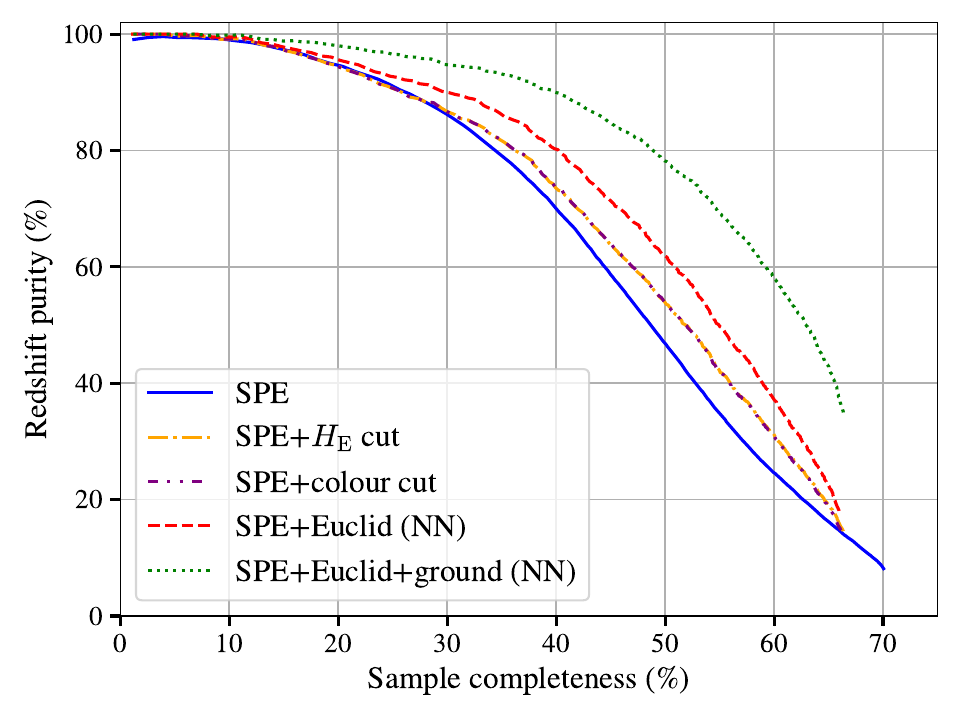}
    \caption{Redshift purity and sample completeness as a function of spectroscopic reliability threshold. The solid blue, dashed red, dotted green, and dash-dotted orange lines respectively correspond to a selection using only SPE reliability, SPE reliability combined with a photometric classification based on \Euclid data,  with the classification that uses \Euclid and ground-based photometry, with the $\HE$ magnitude limit selection, and with the colour selection in the $(\IE - \HE)$-$\HE$ plane}. In all cases the recall of the photometric classification is set to $95\%$.
    \label{fig:punco-cases}
\end{figure}

The spectroscopic measurement pipeline carries out a likelihood analysis using spectral templates to estimate the redshift. It produces a probability distribution function of the redshift that is typically sharply peaked with a few primary redshift solutions. The integral of the peak provides a useful measure of the reliability of the solution. We vary the threshold in this reliability value to select spectroscopic samples and build the relationship between redshift purity and completeness, as shown by the SPE solid blue line in Fig.~\ref{fig:punco-cases}.

In the following discussion we focus on the results from the neural network classifier applied to the simulated spectroscopic sample. The purity and completeness values should be taken as indicative of the general trends and not as accurate forecasts of the pipeline performance. The values depend on the specific distribution of simulated sources and instrumental configuration. The target sample is defined as described in Sect. \ref{sec:labels-pre}, using the total flux of the H$\alpha$ and \ion{N}{II} complex.

\begin{figure}
    \centering
    \includegraphics[width=.5\textwidth]{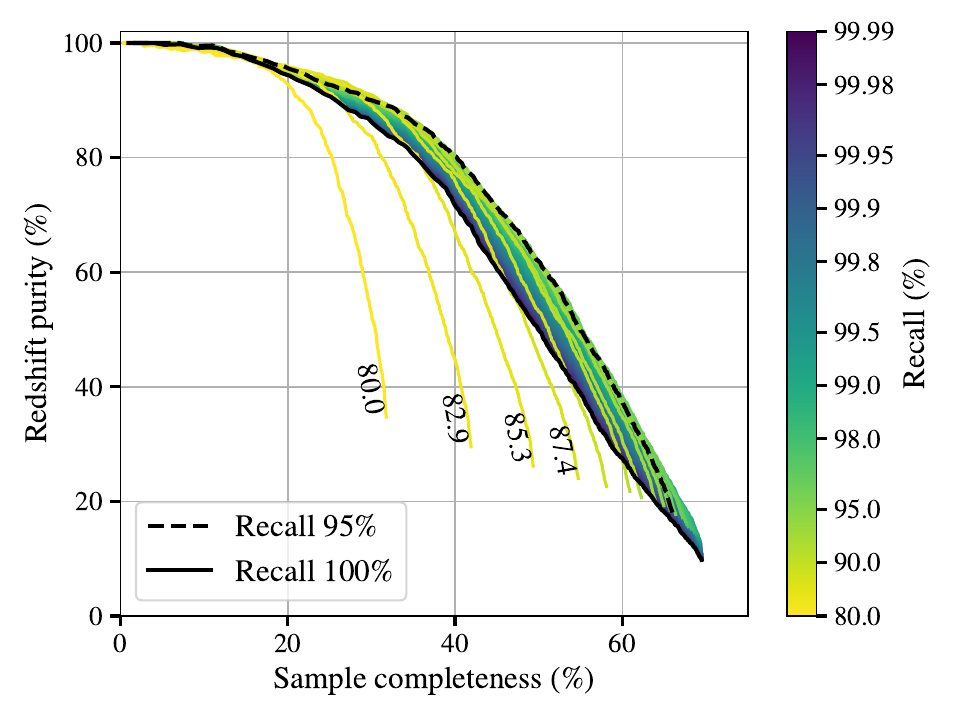}
    \includegraphics[width=.5\textwidth]{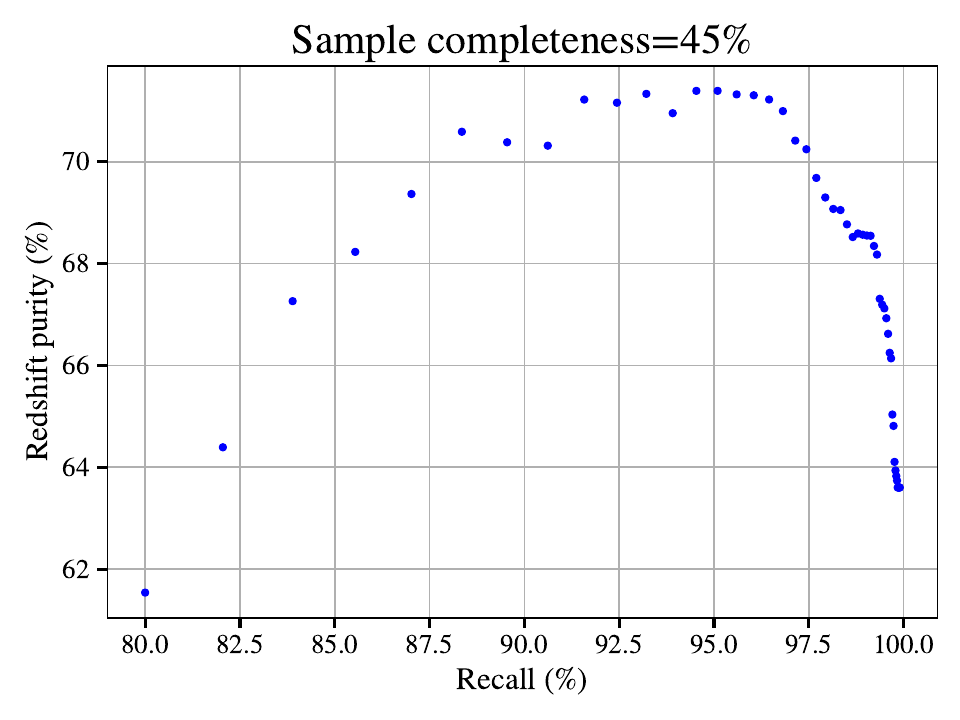}
    \caption{Spectroscopic redshift purity and completeness with the addition of the photometric classification. \emph{Top}: 
    the curves are colour-coded as a function of the recall of the photometric classification. The spectroscopic reliability threshold varies along each curve, while varying the threshold on the photometric classification probability shifts the curve.  The purity improves as recall increases, reaching a maximum for recall ${\sim} \, 95\%$ and declining after. For better visualisation, the first lines are labelled with the corresponding recall value. At recall values above $95\%$ the curves are tightly packed. The solid black line corresponds to $100\%$ recall, while the dashed line to $95\%$ recall, the value we chose to benchmark our results. \emph{Bottom}: redshift purity as a function of the recall of the photometric classification, fixing the value of sample completeness to $45\%$. }
    \label{fig:purity-recall}
\end{figure}

Figure~\ref{fig:punco-cases} shows how redshift purity versus sample completeness plot improves when we complement the pure spectroscopic reliability cut selection (blue solid line) with increasing information provided by the photometric neural network classifier for the two configurations using \Euclid-only or \Euclid and ground-based photometry. The curve corresponding to the $\HE$ magnitude-limit selection that gives $95\%$ recall (see Table~\ref{tab:Hcut95recall}) is also plotted together with the curve corresponding to the colour selection presented in Sect.~\ref{sec:colour-selection-benchmark}. These two curves visually overlap, but the colour cut curve (dash-dot-dotted purple line) is marginally higher than the simple magnitude cut curve (dash-dotted orange line). The figure shows that in the range between $40\%$ and $60\%$ completeness, the photometric classification improves the redshift purity. For example, at a fixed value of $45\%$ sample completeness, the classification based on \Euclid-only bands improves the purity by ${\sim} \, 20\%$, when we add ground-based photometry the improvement rises to ${\sim} \, 45\%$. The simple $\HE$ magnitude limit selection, at that same completeness value, gives an improvement of a few per cent only (${\lesssim} \, 10\%$), evidencing the importance of exploiting all available photometric information. 

To examine the effect of the photometric classification in more detail, in the top panel of Fig.~\ref{fig:purity-recall} we show the redshift purity and sample completeness as a function of the reliability threshold imposed on the spectroscopic redshift measurement. The photometric classification has its own threshold parameter on the classification probability, which when combined with the spectroscopic selection, produces a family of curves. We label these curves based on their recall values. The bottom panel shows the dependence of redshift purity on the photometric selection recall, when the completeness is fixed to $45\%$. 
As we see, a recall value of $95\%$ approximately maximises the purity-completeness curve, which justifies the choice made in Sect. \ref{sec:comparison_metrics}. 

The main conclusion from this exercise is that the impact of properly elaborated photometric information on the final purity and completeness of the \Euclid spectroscopic sample is very significant, with a major improvement especially when ground-based visible bands are included. The precise gain, however, will depend on the galaxy distribution, the survey configuration and the instrument model.

\section{Conclusions}\label{sec:conclusions}

We have investigated the benefits of combining photometric information with the spectroscopic measurement criteria for selecting \Euclid spectroscopic samples. \Euclid spectroscopy will give estimates of the galaxy redshifts, fluxes of the emission lines, and confidence intervals. However, since emission-line galaxies make up only a small fraction of the photometric sample, measurement noise can reduce the redshift purity and completeness of the sample and degrade the figure of merit for the galaxy clustering probe.
The addition of photometric criteria in the selection can allow us to improve the purity of the sample by identifying sources that are likely to be bright emission-line galaxies at the target redshift.

To this end, we compared a set of machine learning classification algorithms with the aim of photometrically selecting emission-line target galaxies that are likely to give good redshift measurements in the Euclid Wide Survey. We used two catalogues to benchmark the classification performance, EL-COSMOS and Flagship2. Both catalogues have \Euclid and ground-based simulated photometry. We produced noisy realisations of the catalogues assuming background-limited observations. The two catalogues yield similar results when using as input \Euclid-only photometry, but when this is combined with ground-based data, the results using Flagship2 outperform those with EL-COSMOS. This is related to the differences in the H$\alpha$ luminosity function and colour distribution of the two catalogues.
In addition to these two configurations (\Euclid-only and \Euclid plus ground) we also considered adding morphological information (half-light radius and the axial ratio). We find that in general, while the addition of ground-based data strongly improves the precision (doubling it in the case of Flagship2), including morphological information (at least in the form provided here) gives negligible improvement.

The purity of the final spectroscopic sample will depend on the combination of the photometric classification with further selection criteria based on the properties of the spectroscopic data (see diagram in Fig.~\ref{fig:flowchart}).
To investigate this requires full end-to-end simulations of the spectroscopic reduction pipeline. We presented a preliminary exercise to assess the relative gain when the spectroscopic data are complemented by the photometric selection discussed here. This will be expanded in future work. We showed that in the range between $40\%$ and $60\%$ completeness the purity is boosted by ${\sim} \, 20\%$ when using \Euclid-only bands, and between $40\%$ and $100\%$ when including ground-based photometry.  We consider this a remarkable indication.

The introduction of ground-based data significantly improves the purity of the sample, but in the practical application can also bring additional nuisance in the form of systematic errors. The ground-based photometry will come from multiple surveys and so will not be fully homogeneous. It will also suffer from additional selection effects correlated with the observing conditions that can propagate as systematic errors to the galaxy clustering measurements and cosmological constraints. Thus, the gains in purity from incorporating ground-based data must be carefully weighed against the potential of adding systematic errors, also considering the specific requirements of the science analysis to be carried out. We foresee that ground-based data may be used in analyses where a higher level of purity is desired, such as for studying the galaxy halo occupation distribution or galaxy evolution as a function of environment.

Photometric redshifts can also play a key role in sample selection. We used the \Euclid photometric redshift estimates to select galaxies in the target redshift range and compared the performance of such a selection to that of the colour-based machine learning classifiers. Figure~\ref{fig:FPR} shows that the photo-$z$ selection is very efficient for redshift classification, especially to remove low redshift interlopers, but is not effective in identifying emission-line galaxies. Indeed, the photo-$z$ selection has the highest fraction of false positives from faint galaxies with $f_{\rm H\alpha+\left[\ion{N}{ii}\right]} < \num{2e-16}  \, {\rm erg} \, {\rm s}^{-1} \, {\rm cm}^{-2}$, but the lowest for bright ones with $f_{\rm H\alpha+\left[\ion{N}{ii}\right]} > \num{2e-16}  \, {\rm erg} \, {\rm s}^{-1} \, {\rm cm}^{-2}$, which means that it makes a better redshift selection than the algorithms presented in this work. Photometric redshifts could be used with additional constraints from spectral energy distribution fits to identify bright emission-line galaxy targets. In particular, the \Euclid photometric redshift pipeline will output estimates of galaxy physical properties including the star-formation rate and dust attenuation, which will allow us to select emission-line galaxy samples. We expect that a classifier developed based on photometric redshifts and estimates of physical properties from spectral energy distribution fitting would perform similarly to the pure colour and magnitude-based classifiers that we tested, since the underlying photometric information is the same. Analogously, we expect a classifier trained to make a selection in redshift alone to perform similarly to the photo-$z$ selection. Alternative classifiers that use the estimates of galaxy physical properties from the \Euclid photometric redshift pipeline for sample selection will be investigated in a future work.

It is important to note that in this study some of the complications that will be present in real \Euclid data were not considered. First, we assume an ideal training set, which is fully representative of the Wide Survey data. In the actual Euclid Wide Survey, the training set will come from the Deep Fields, which will total $\sim 50 \, \text{deg}^2$. Shallow and full-depth photometric measurements will be available for the \Euclid photometry in the Deep Fields; however, we will only have the full-depth measurements for the ground-based photometry. As they are currently trained, the machine learning algorithms learn to classify the targets at a given noise level and it is not necessarily true that they will be able to generalise their results when trained and tested on samples with different noise levels. Therefore, if ground-based photometry is used, it will be necessary to degrade the measurements to match the noise level in the Wide Survey. Since the ground-based photometry will come from multiple surveys, this operation will not be simple, and residual variations in homogeneity in the noise can lead to systematic variations in the classifier performance.

Moreover, the effective emission-line flux limit will vary across the Wide Survey due to foreground emission including zodiacal light and scattered stellar light \citep{EuclidWide}. In this study, we used a fixed flux limit to build the training set of emission-line galaxies. In practice, this does not impose a sharp flux cut in the measured sample. However, when developing a classifier on real data, we will be able to use the Deep Survey to define the training set as the set of galaxies that are correctly measured by the \Euclid pipeline, without imposing any specific constraints on their physical properties. It will also be possible to construct a classifier that accounts for variations in the noise level across the Wide Survey to optimise the sample. 

Finally, a further complication that must be considered is contamination from stars in the galaxy catalogue that can impact the purity. The photometric classifier can be trained to maximise the precision in the presence of stars. This work will require us to incorporate a star-galaxy classifier, which is based both on size and photometric colours. Here, the morphological measurements will be important.

In the next stage of this work, we will consider the full set of spectroscopic and photometric selection criteria in order to compute the redshift purity, sample completeness and ultimately cosmologically relevant figures of merit. This requires running the spectroscopic reduction pipeline on mock data in order to produce end-to-end simulations. Such simulations will allow us to optimise the sample selection criteria, possibly with the use of machine learning classifiers. 
With \Euclid observations beginning in fall 2023, we will be able to further tune the selection based on the actual telescope performance and ultimately construct the spectroscopic galaxy sample that will be used to test the cosmological model.

\begin{acknowledgements}
    \AckEC
            
    This work has made use of CosmoHub \citep{2017ehep.confE.488C,TALLADA2020100391}.

    CosmoHub has been developed by the Port d'Informació Científica (PIC), maintained through a collaboration of the Institut de Física d'Altes Energies (IFAE) and the Centro de Investigaciones Energéticas, Medioambientales y Tecnológicas (CIEMAT) and the Institute of Space Sciences (CSIC \& IEEC), and was partially funded by the "Plan Estatal de Investigación Científica y Técnica y de Innovación" program of the Spanish government.
\end{acknowledgements}

\bibliographystyle{aa_url}
\bibliography{euclid_photo_selection}

\begin{appendix}

\section{Colour-magnitude projection planes}\label{sec:col-proj}

\begin{figure}
    \centering
    \includegraphics[width=.49\textwidth]{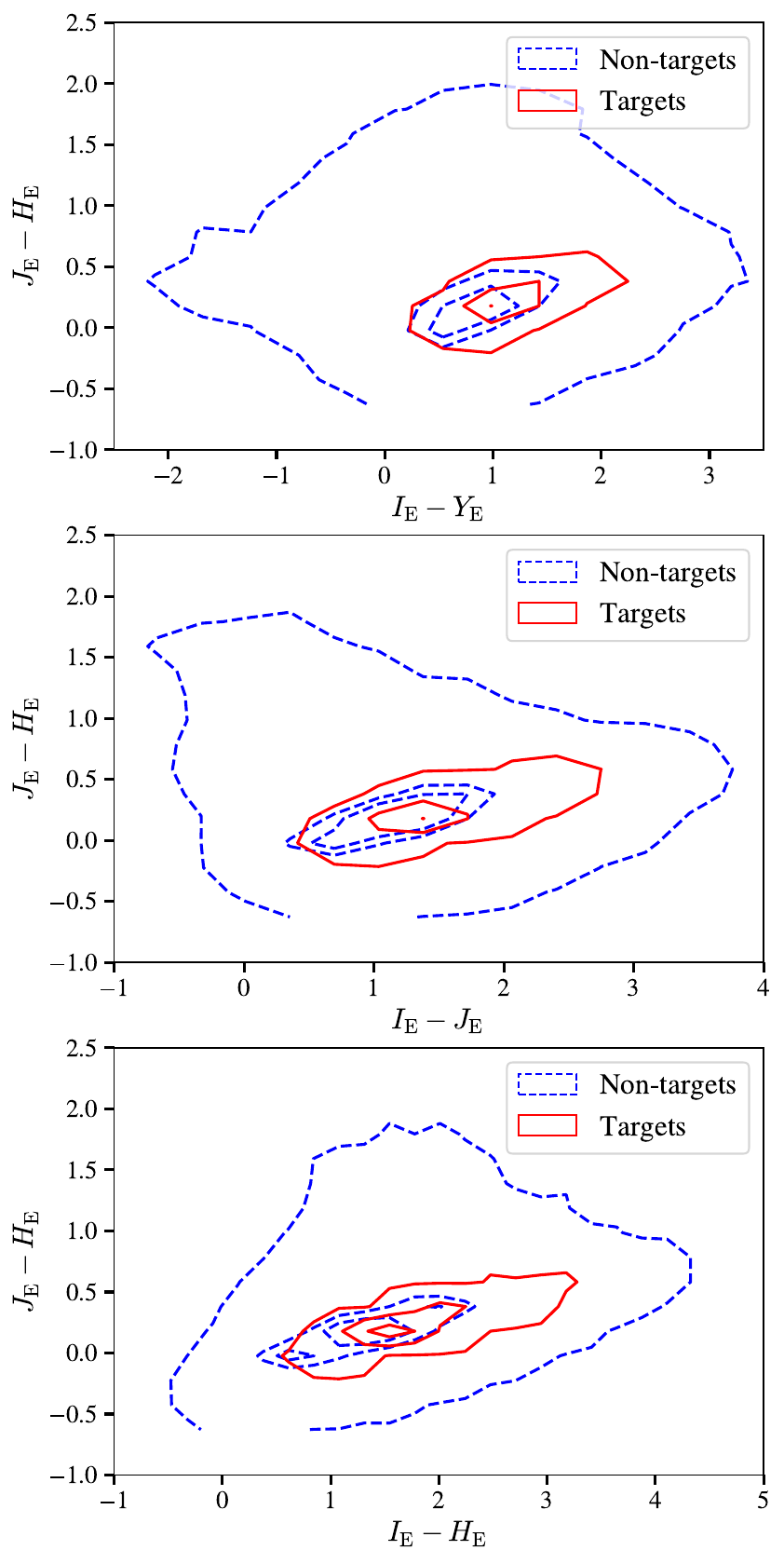}
    \caption{Target and non-target distributions in colour-colour and colour-magnitudes planes for the Flagship2 catalogue. The contours contain $99\%$, $50\%$, and $25\%$ of the samples.}
    \label{fig:color-color}
\end{figure}

We present in Fig.~\ref{fig:color-color} the colour-colour and colour-magnitude distributions for targets and non-targets in the Flagship2 catalogue for three different combinations. The contours contain $99\%$, $50\%$, and $25\%$ of the samples. There is a nearly complete overlap of the targets and non-targets in the colours. 

\section{Photo-$z$ as input variables} \label{app:photz-feat}

\begin{figure}
    \centering
    \includegraphics[width=.5\textwidth]{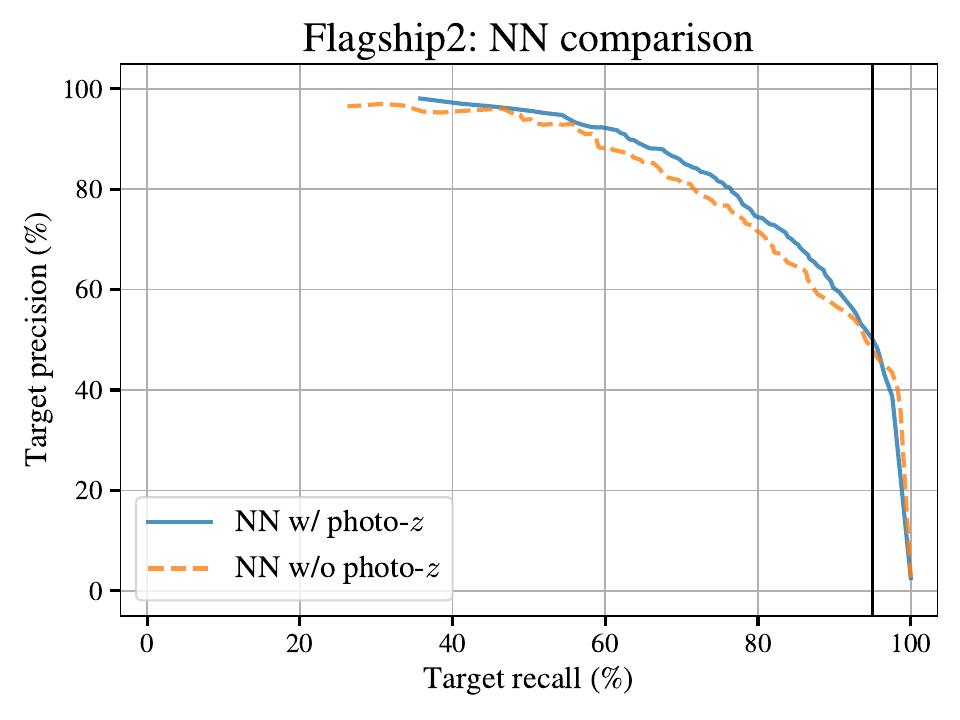}
    \caption{Comparison of the precision versus recall curves of two neural networks trained with and without photo-$z$s as an input feature. The two neural networks were trained with \Euclid and ground-based photometry, but in the case of the solid blue line the algorithm takes the photo-$z$ of the galaxy as an additional feature. The solid vertical line corresponds to $95\%$ recall.}
    \label{fig:NN-comparison}
\end{figure}

As an additional test, we trained a neural network with Flagship2 data in the \Euclid plus ground-based configuration with the additional information of the measured photo-$z$. In principle, the neural network can extrapolate the redshift from the photometric information. However, by directly providing the photo-$z$ we may facilitate the selection process as the network will be able to put more attention on the emission line flux.

The precision at $95\%$ recall is $47.9\%$ in the case without photo-$z$, as reported in Table~\ref{tab:precision95recall}. When we add the photo-$z$ of the galaxy as an input feature the precision rises to $50.1\%$. Nevertheless, the addition of the photo-$z$ to the input information makes the classifier dependent on the complex process used to produce the photo-$z$s, including the training sets, spectral energy distribution models, and algorithms used. We postpone to a future work the detailed study of these dependencies and the performance on realistic data.

\section{Selection probability maps}
\label{sec:probmaps}
    \begin{figure*}
        \centering
        \includegraphics[width=1\textwidth]{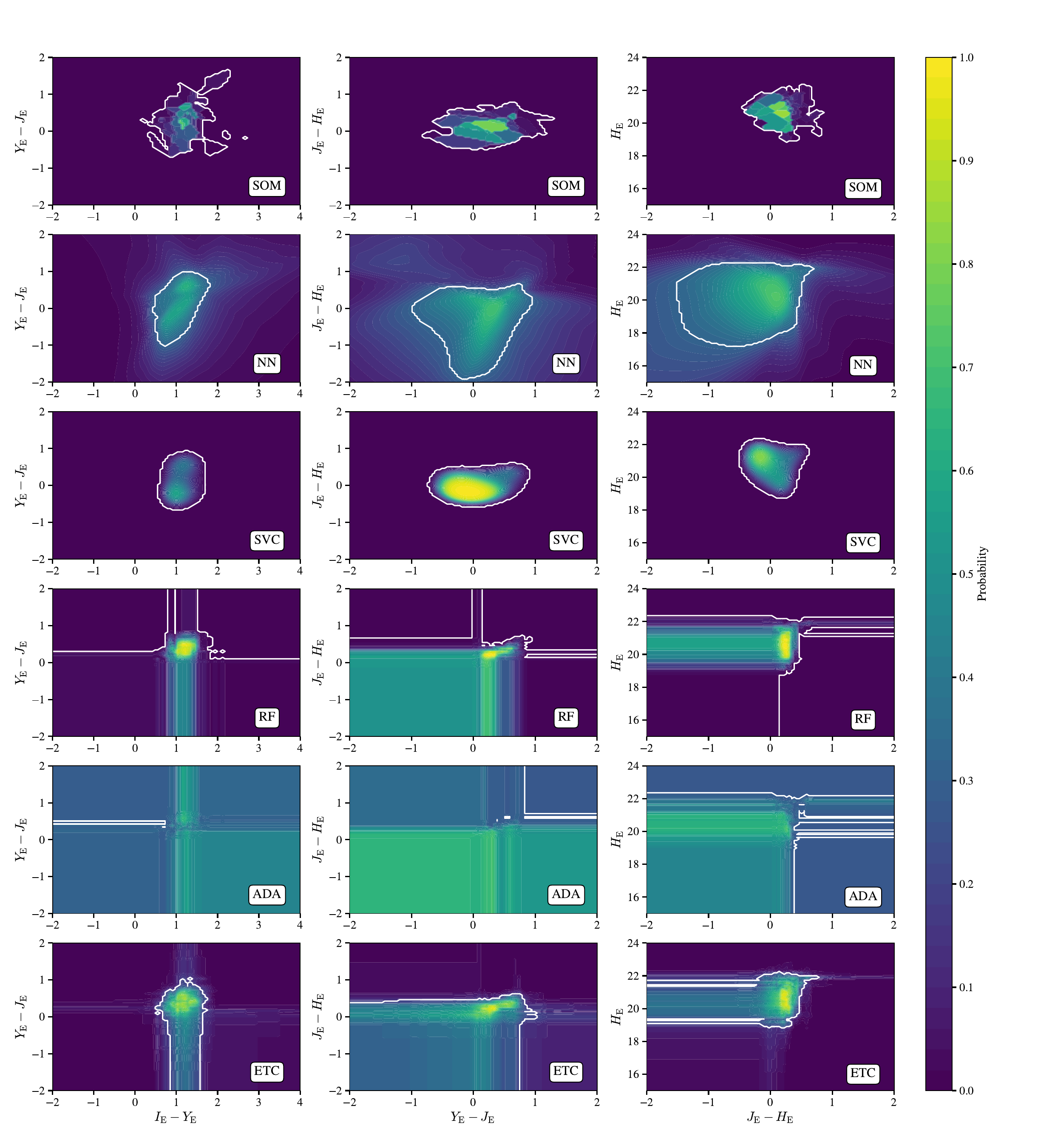}
        \caption{Probability maps in colour-colour and colour-magnitude planes, for the six classifiers tested in this paper, trained using Flagship2 \Euclid photometry only. The thick white contour marks the probability threshold that gives 95\% recall.}
        \label{fig:comp_maps}
    \end{figure*}

Figure~\ref{fig:comp_maps} gives a visualisation of the selection probability for each classifier in planes through the parameter space.  We show the results from the Flagship2 catalogue for the case when classifiers are trained with \Euclid photometry alone. Each row shows the colour-colour and colour-magnitude plots for a given algorithm. The parameter space is four-dimensional, and the two-dimensional planes are made by fixing two of the parameters to their median values.

One notices that the different classifiers identify a similar region of maximum probability for a given pair of features. The shape and the gradients of these regions, however, vary for each algorithm. This is due to the differences in the selection algorithms and possible projection effect when the boundaries are represented on the planes. In the case of the single classifiers (top three rows: self-organising map, neural network, and support vector classifier) they are compact and well defined, unlike for the cases of voting classifiers based on decision trees (bottom three rows). Also, the contours and gradients are less smooth for the self-organising map than for the neural network and the support vector classifier. The probability gradient of the support vector classifier is very steep, especially in comparison to the neural network. 

The probability maps for the voting classifiers (bottom three rows) show orthogonal contours. This is due to the common base classifier of these algorithms, the decision tree, which tends to produce decision rules orthogonal to one another. At the same time, the three algorithms have very different probability contours. These differences are related to the batch selection rule used to train the decision trees (see Sect. \ref{sec:algorithms}). We expect that the algorithms that give a classification model with irregular and steep contours (such as the self-organising map) or stepped contours (such as the decision trees) will be prone to over-fitting and will show poorer performance than algorithms that give smooth probability contours.
        
\end{appendix}

\end{document}